\documentclass[a4paper,fleqn,usenatbib]{mnras}

\usepackage{txfonts}

\usepackage[T1]{fontenc}
\usepackage{ae,aecompl}
\usepackage{substitutefont}
\substitutefont{TS1}{aer}{cmr}

\usepackage{graphicx}	
\usepackage{amssymb}	
\makeatletter
\renewcommand{\fnum@figure}{Fig. \thefigure}
\makeatother
\usepackage[autostyle]{csquotes}

\title[Kinematically anomalous HI gas]{\bf A new technique to isolate kinematically anomalous gas in HI data cubes} 

\author[N. Randriamiarinarivo et al.]{
	N. Randriamiarinarivo,$^{1}$\thanks{E-mail: nandrianiana@gmail.com (NR)}
	E. C. Elson,$^{1}$
	A. J. Baker$^{1,2}$
	\\
	$^{1}$Department of Physics and Astronomy, University of the Western Cape, Robert Sobukwe Rd, Bellville, 7535, South Africa\\
	$^{2}$Department of Physics and Astronomy, Rutgers, The State University of New Jersey, 136 Frelinghuysen Road, Piscataway, NJ 08854-8019, USA
}

\date{Accepted 18 November 2022. Received 28 June 2022; in original form 30 November 2021}

\begin{document}

\label{firstpage}
\pagerange{\pageref{firstpage}--\pageref{lastpage}}
\maketitle
	
\begin{abstract}
H\,{\sc{i}} line observations of nearby galaxies often reveal the presence of extraplanar and/or kinematically anomalous gas that deviates from the general circular flow. In this work, we study the dependence of kinematically anomalous H\,{\sc {i}} gas in galaxies taken from the {\sc Simba} cosmological simulation on galaxy properties such as H\,{\sc {i}} mass fraction, specific star formation rate, and local environmental density. To identify kinematically anomalous gas, we use a simple yet effective decomposition method to separate it from regularly-rotating gas in the galactic disk; this method is well-suited for application to observational datasets but has been validated here using the simulation. We find that at fixed atomic gas mass fraction, the anomalous gas fraction increases with the specific star formation rate. We also find that the anomalous gas fraction does not have a significant dependence on a galaxy's environment. Our decomposition method has the potential to yield useful insights from future H\,{\sc {i}} surveys.

\end{abstract}

	
	
	
\section{Introduction} \label{s-intro}
	
Neutral atomic hydrogen (H\,{\sc{i}}) represents a key phase in the \enquote{baryon cycle} of accretion, star formation, and outflow in galaxies, in part because it represents a large fraction of a typical galaxy's interstellar medium (ISM) by mass, and in part because it is more easily observable than many other ISM (and
circumgalactic medium, CGM) phases. A galaxy's H\,{\sc{i}} content reflects its recent history of accretion from the intergalactic medium (IGM) and indirectly supplies its main fuel for star formation; it is therefore critically linked to its overall evolution (e.g., \citealt{Kannappan_2013}). While H\,{\sc{i}}
is most frequently found in rotationally dominated gas disks \citep{1981AJ.....86.1791B, 10.1093/mnras/sty275}, these can extend well beyond galaxies' stellar disks and exhibit significant irregularities (asymmetries, warps, tails, etc.) in their outer reaches. With new surveys now actively observing H\,{\sc{i}} in large samples of galaxies spanning different combinations of area and depth (e.g., \citealt{Blyth:2018RG}), it is important to find reliable methods of extracting information from galaxies' H\,{\sc{i}} distributions and kinematics, and using this information (in conjunction with observations at other wavelengths) to shed light on the processes governing mass assembly and draw connections to the other important
drivers of galaxy evolution. In particular, isolating and characterizing the neutral atomic gas that is {\it not} confined to a galaxy's disk holds a great deal of promise as a means of understanding its evolutionary trajectory.

\cite{Swaters_1997} and \cite{Fraternali_2001} find that a fraction of H\,{\sc {i}} in some galaxies is situated in a thick layer that can extend vertically out to a few kpc above the midplane of the galactic disk. This gas has a typical scale height of $1 - 3\,{\rm kpc}$ and typically lags the main H\,{\sc {i}} disk in terms of rotation speed \citep{2007AJ....134.1019O, 2000A&A...356L..49S, 2011ApJ...740...35Z, 2019A&A...631A..50M, 2001ApJ...562L..47F, 2005ASPC..331..239F}.
Such extraplanar gas (EPG) has also been observed in some galaxies to exhibit non-circular motion ~\citep{2005A&A...439..947B, 2002AJ....123.3124F, 2019A&A...631A..50M}, and in some systems to be in the inner part of the disk ~\citep{2002AJ....123.3124F, 2008A&A...490..555B}.
\citet{2007AJ....134.1019O} and \citet{2019A&A...631A..50M} find that EPG is most frequently found and most prominent in late-type galaxies, where it comprises 10 to 30\% of the gas in the disk. \cite{2019A&A...631A..50M} highlight and discuss the properties of EPG in disk galaxies, and find that it decreases in rotation velocity at a rate of approximately 10\,${\rm km\ s}^{-1}$ ${\rm kpc}^{-1}$ above the midplane. They also measure an inflow velocity of approximately $20 - 30\,{\rm km\ s}^{-1}$ towards the inner disk. 
Studies of ionized EPG have also been conducted in various galaxies, including the Milky Way \citep{2009RvMP...81..969H}. It has been shown that just like the EPG observed in H\,{\sc {i}}, ionized EPG also exhibits lagging rotation \citep{2007A&A...468..951K} and non-circular motion \citep{2004A&A...424..485F}. \citet{2003A&A...406..493R} and \citet{2003ApJS..148..383M} find that ionized EPG is most prevalent in late-type galaxies. We note that the properties of such ``classical'' EPG may not apply to other gas reservoirs in the vicinity of galaxies --- e.g., material that is inflowing from the CGM or being accreted through minor mergers --- whose kinematics can be completely decoupled from those of the gas in the disk.
Numerical simulation is a very powerful tool for understanding the physics of gas in galaxies and how it interacts with the ambient medium. \cite{2011MNRAS.415.1534M} study the interaction between cold clouds ejected from a galactic disk and the corona of the disk by using two-dimensional hydrodynamical simulations of cloud-corona interactions, in work later extended by \cite{10.1093/mnras/stw1930}. 
In this work, we use the {\sc Simba} suite of cosmological galaxy formation simulations \citep{Dave:2019yyq}, which is the descendent of the MUFASA simulation \citep{Dav__2016}, together with a new method of decomposing the H\,{\sc{i}} content of galaxies into kinematically regular and anomalous components.
Our main goal in developing this method is to provide a tool that can be identically applied to both simulations and observations of galaxies. To validate it, we here analyze the properties of the {\sc Simba} particles in a way that cannot be done for observations. The layout of this paper is as follows. In Section ~\ref{s-sims}, we present the samples of {\sc Simba} galaxies we use in our study (by design, these are rotation-supported disks at $z = 0$). Section \ref{s-method} describes our new method of decomposing a galaxy's emission into kinematically regular and anomalous components, including its validation using a particle-based analysis of the simulation and a discussion of the effects of inclination. 
\section{Simulated galaxies} 
\label{s-sims}
	
{\sc Simba} uses the {\sc Gizmo} multi-simulation code \citep{Hopkins_2015} in its Meshless Finite-Mass (MFM) mode, with multiple comoving volumes, and extends from redshift $z\ = 249$ to $z\ = 0$. 
The reference {\sc Simba} simulation includes $1024^3$ gas and $1024^3$ dark
matter particles (with particle masses $1.82 \times 10^7\,M_\odot$ and $9.6 \times 10^7\,M_\odot$, respectively) in a comoving volume of $100^3\,h^{-3}\,{\rm Mpc}^{-3}$.  In this paper, however, we use an alternative simulation that is higher in mass resolution by a factor of 8, containing $512^3$ gas and $512^3$ dark matter particles in a $25^3\,h^{-3}\,{\rm Mpc}^{-3}$ comoving volume.
We use the CAESAR package \footnote{https://caesar.readthedocs.io/en/latest/} to crossmatch galaxies to their haloes, and to extract important properties of the galaxies such as star formation rates (SFRs), stellar masses, and H\,{\sc {i}} masses. 

In this work, our mock data cubes are generated directly from the particles in the simulated galaxies. We select galaxies at $z = 0$ with stellar mass $M_\star > 10^8\,M_\odot$ to avoid dwarf galaxies. For each gas particle, we compute its H\,{\sc {i}} content by following \citet{2009A&A...504...15P}, who define H\,{\sc {i}} fraction $f_{\rm HI}$ as
\begin{equation}
f_{\rm H\,I} = \frac{2C+1 - \sqrt{(2C+1)^2 - 4C^2}}{2C}
\end{equation}
with
\begin{equation}
C = \frac{n\beta(T)}{\Gamma_{\rm HI}}
\end{equation}
where $n$, $T$, and $\Gamma_{\rm H\,I}$ are the hydrogen density, gas temperature, and H\,{\sc{i}} photoionization rate, respectively. $\beta(T)$ is the recombination rate coefficient, which follows an analytic expression described by ~\citet{1996ApJS..103..467V} as
\begin{equation}
\beta(T) =  a\left[\sqrt{T/T_0}(1 + \sqrt{T/T_0})^{1-b} (1 + \sqrt{T/T_0})^{1+b}\right ]^{-1}
\end{equation}
 For H\,{\sc{i}}, $a = 7.982\times 10^{-11}\,{\rm cm^3\,s^{-1}}$, $b = 0.7480$, $T_0 = 3.148\,{\rm K}$, and $a = 7.036\times 10^5\,{\rm K}$. For $\Gamma_{\rm H\,I}$, we follow ~\citet{2009A&A...504...15P} in using the photoionization background of ~\citet{2001cghr.confE..64H}, i.e., $\Gamma_{\rm H\,I}$ = $10^{-13}\,{\rm s^{-1}}$ at $z = 0$.

\begin{figure*}
		\includegraphics[scale = 0.33]{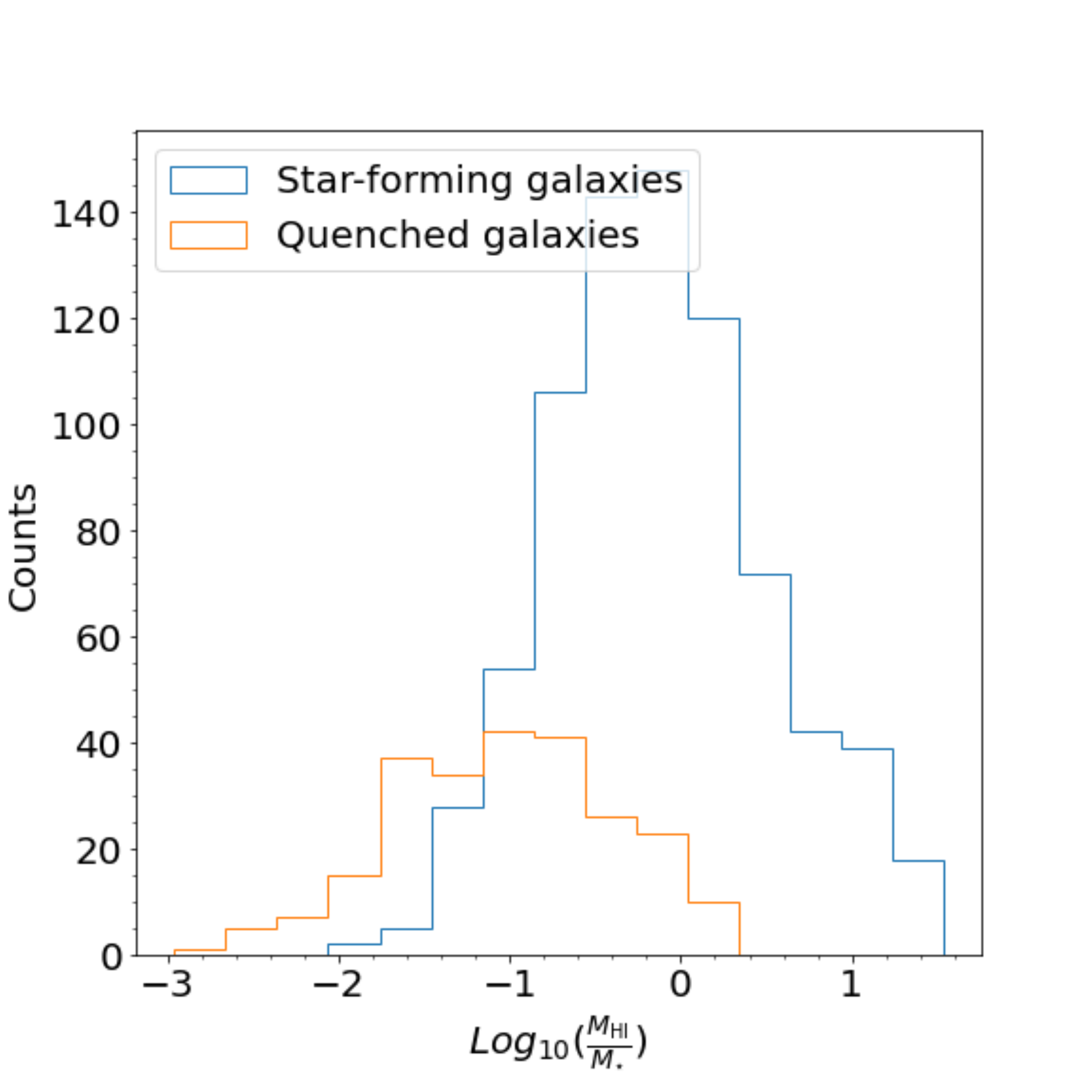}	
		\includegraphics[scale = 0.33]{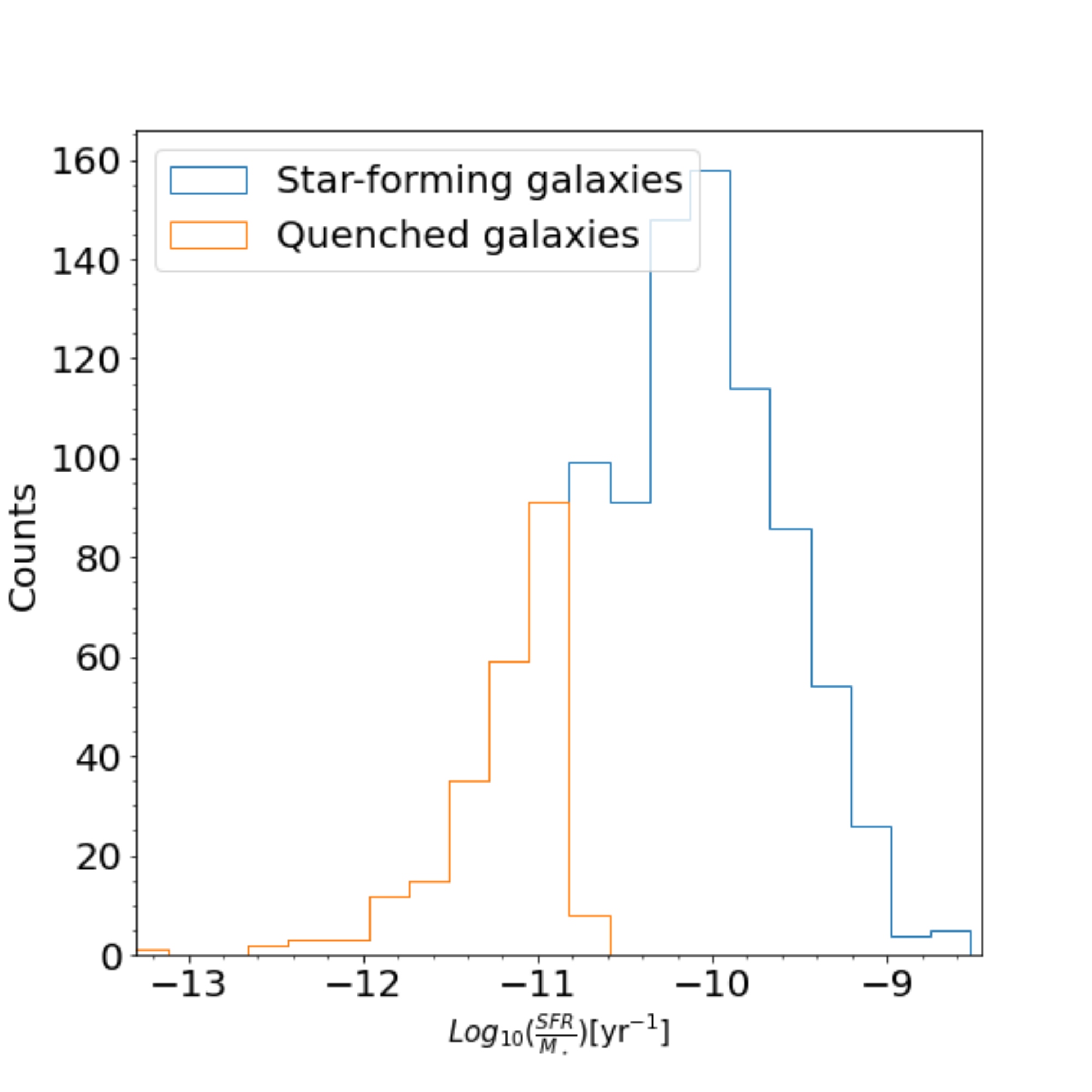}
		\caption{Distribution of H\,{\sc {i}} gas mass fraction (left) and specific star formation rate (right). Orange and blue curves represent star-forming and quenched galaxies, respectively.}
		\label{fig:distprop}
	\end{figure*}

In addition to the mass threshold noted above, we select galaxies that are not bulge-dominated using the $\kappa_{\rm rot}$ parameter, which is the fraction of kinetic energy invested in ordered rotation and is defined as
\begin{equation}
\kappa_{\rm rot} \equiv \frac{K_{\rm rot}}{K}
\end{equation}
where
\begin{equation}
K_{\rm rot} \equiv \sum_i \frac{1}{2} m_i \left( \frac{{j_s}_i}{R_i} \right )^2
\end{equation}
the sum is over all the particles in a galaxy, $j_s$ is the specific angular momentum of a single particle, and $K$ is the total kinetic energy of all particles in a galaxy.
$\kappa_{\rm rot} < 0.5$ indicates a spheroid-dominated galaxy, while $\kappa_{\rm rot} > 0.7$ refers to disk-dominated galaxy, and intermediate values of $\kappa_{\rm rot}$ mark galaxies that have comparable dispersion and rotational velocity ~\citep{20975}. In this study, we choose $\kappa_{\rm rot} > 0.6$ to ensure that our sample includes ``disky'' galaxies with a range of intermediate types, in order to match a more realistic observed sample.
In order to simulate an observation with a facility like MeerKAT, we construct cubes with $5\,{\rm km\,s^{-1}}$ velocity channels and a spatial resolution of 10\,arcsec. We smooth the cubes spatially and spectrally by effective convolution with a Gaussian kernel, to resolutions of full width half power 30\,arcsec and $15\,{\rm km\,s^{-1}}$, respectively. For our reference orientation, we assign each galaxy to have inclination angle $\theta = 60^\circ$.  
For this study, we also separate our selected sample into star-forming galaxies and quenched galaxies by applying the specific star formation rate cut-off $\rm {log(sSFR) [Gyr^{-1}]} > - 1.8 + 0.3$ $z$ ~\citep{2019MNRAS.486.2827D}. In Fig.~\ref{fig:distprop}, we present the distributions of atomic gas fraction and specific star formation rate for the 303 galaxies in our sample, in the left and right panels respectively.

\section{Methods} \label{s-method}

\subsection{A cube-based approach to identifying anomalous gas}

\begin{figure*}
		\centering \includegraphics[scale = 0.3]{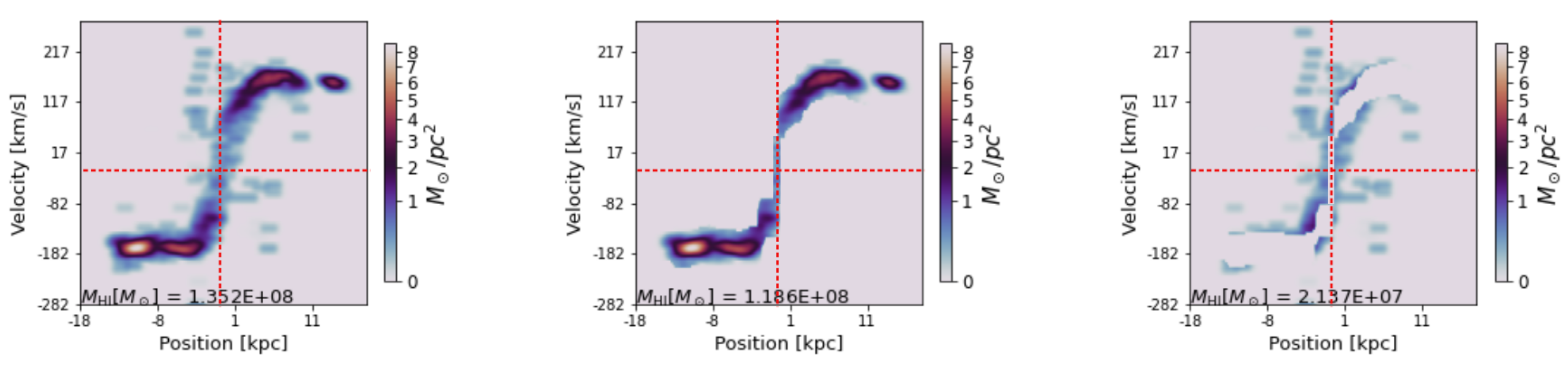}
		\caption{Position-velocity (P-V) plots of a galaxy from the simulation. The x and y axis values are subtracted by the position of the centre of mass of the galaxy and its systemic velocity, respectively. The left panel shows the P-V diagram for the full cube containing all of the galaxy's H\,{\sc {i}} emission. The black circles indicate some of the most conspicuous examples of kinematically anomalous material, which is the sort of gas that our decomposition method has been designed to identify and extract. The middle panel shows the gas that comes from the main disk of the galaxy; the right panel shows the anomalous gas that has been extracted from the original cube. Values at the bottom indicate the total H\,{\sc {i}} mass present in the respective P-V slices. The vertical and horizontal red dashed lines represent the position and velocity of the centre of mass the galaxy.}
		\label{fig:PV1}
	\end{figure*}

For our sample of simulated galaxies, we aim to decompose H\,{\sc {i}} data cubes into contributions from regularly-rotating H\,{\sc {i}} and kinematically anomalous H\,{\sc {i}}. Given an H\,{\sc {i}} data cube, a position-velocity (P-V) slice extracted along the major axis of the galaxy allows for the regularly-rotating H\,{\sc {i}} component to be easily identified (the P-V slices have a spatial width of 1 pixel = 10\,arcsec). In Fig.~\ref{fig:PV1}, we present the P-V diagram for a galaxy from the simulation at redshift $z = 0$ that has a total H\,{\sc {i}} mass of approximately $9.5\times 10^9 M_\odot$. The regularly-rotating H\,{\sc {i}} component is clearly visible as the very bright S-shaped emission in the P-V slice. The kinematically anomalous component (some parts of which are circled in Fig.~\ref{fig:PV1}) is revealed as the fainter emission that is vertically offset from the regular component. 

\begin{figure*}
		\centering \includegraphics[scale = 0.38]{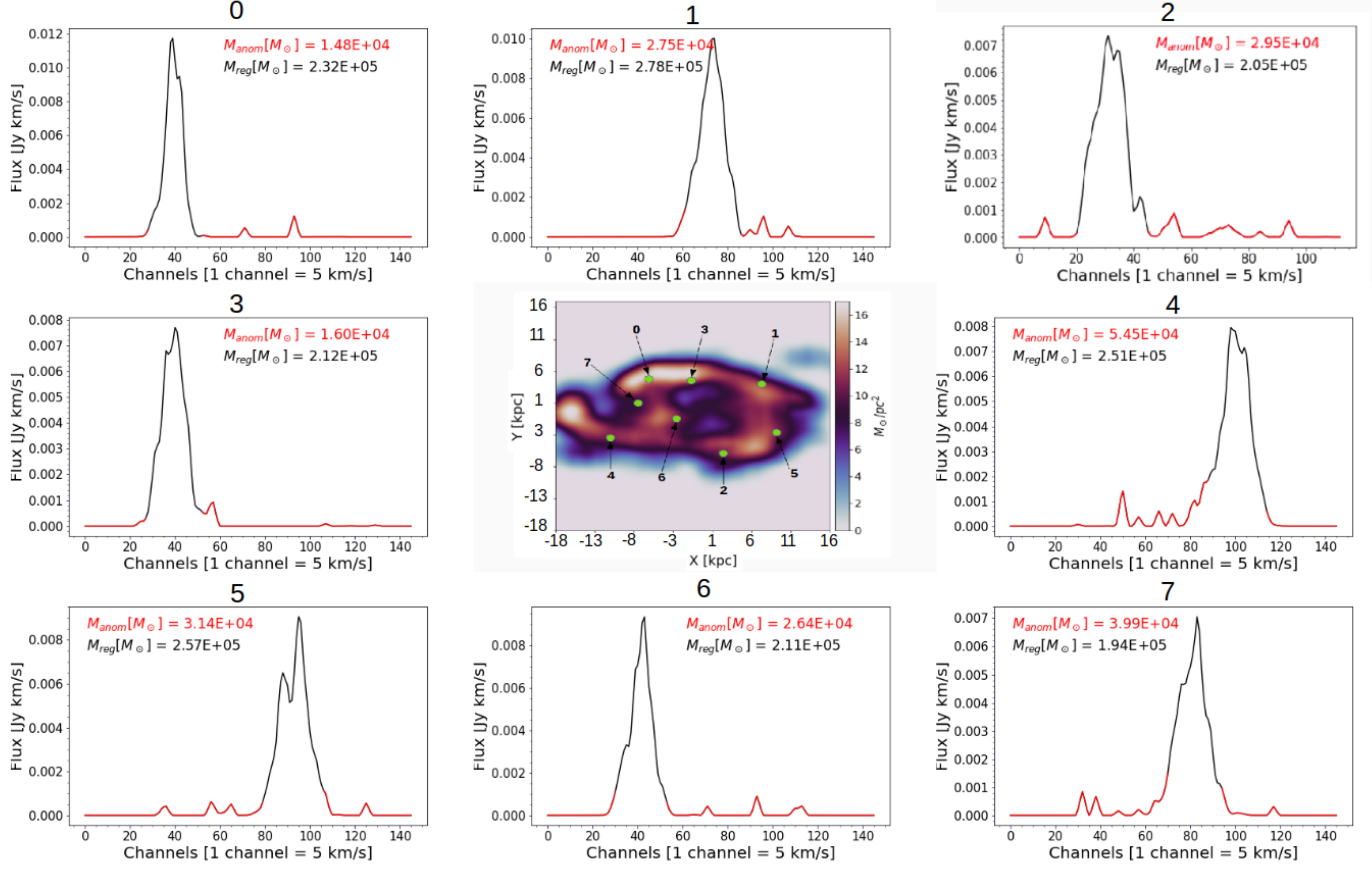}
		\caption{H\,{\sc {i}} moment 0 map of a galaxy from the simulation (center panel) alongside line profiles extracted at the positions numbered on the H\,{\sc {i}} map. The red lines in the profiles denote the emission from the galaxy's anomalous gas, while the black lines represent regular emission.}
		\label{fig:methods}
	\end{figure*}

The first panel of Fig.~\ref{fig:PV1} shows the P-V diagram of the full H\,{\sc {i}} data cube for this galaxy. 
To separate the emission from the regularly-rotating gas and the kinematically anomalous gas in our galaxies, we use H\,{\sc {i}} profiles extracted at individual spatial pixels. The central panel of Fig.~\ref{fig:methods} shows the total intensity map for a simulated H\,{\sc {i}} data cube. At the spatial positions indicated by the numbers 0--7, we have extracted the line profiles along the spectral axis of the cube. These profiles are shown in the surrounding panels. 
Our decomposition method aims to separate the dominant component from the other components. 

\begin{figure*}
		 \centering \includegraphics[scale = 0.19]{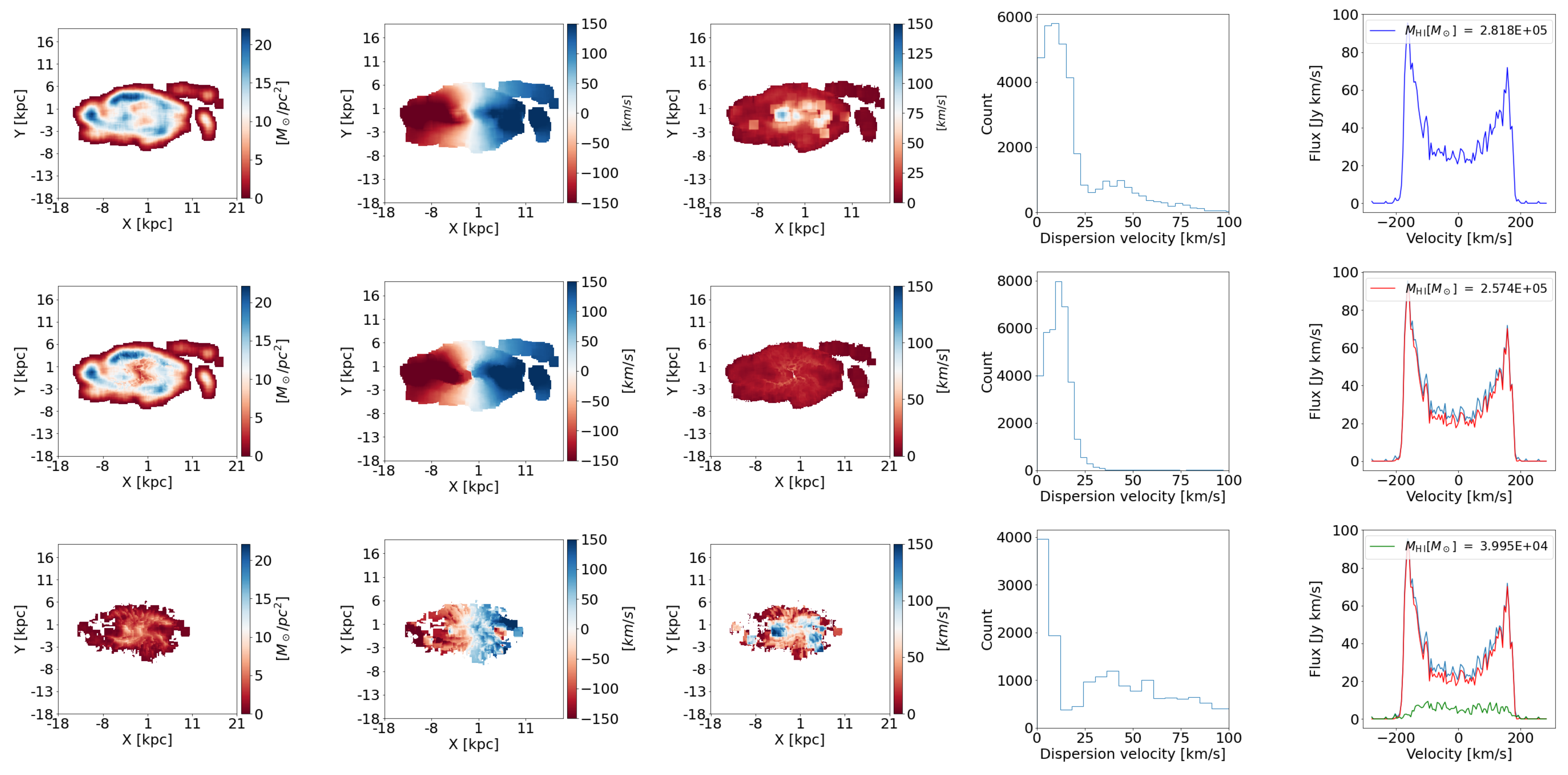} 
		\caption{Dissection of H\,{\sc {i}} in a galaxy from the simulation at $z = 0$. The first through third columns represent the moment 0, 1, and 2 maps, respectively, and the fourth and fifth columns show the distributions  of velocity dispersion and the H\,{\sc {i}} profiles. The first row shows the original simulated data cube, while the second and third rows show the regular and anomalous gas, respectively. }
		\label{fig:panel}
	\end{figure*}

We identify as regular emission all channels that have flux densities within 20\% of the peak flux density of a given line profile. We find that value to be the optimal choice for including all of the main peak emission of the galaxy. For each of these channels, we further identify channels within $\pm 30\,{\rm km\ s}^{-1}$ and include them as regular emission. We choose $\pm 30\,{\rm km\ s}^{-1}$ as the optimal velocity window size based on experiments in which we used different window sizes to try to isolate the regular emission in the line profiles of the galaxies. Any channels that are not within 30\,${\rm km\,s^{-1}}$ of a channel with peak flux at least 80\% as high as the peak flux of the entire line profile are taken to contain kinematically anomalous emission. 
The red and black portions of the line profiles shown in Fig.~\ref{fig:methods} represent the anomalous and regular kinematic components, respectively.
For each of the profiles that we show here, we indicate in the upper right of each panel the amounts of anomalous and regular gas detected at the given position.   
Fig.~\ref{fig:panel} shows the results of our decomposition method applied to the galaxy from Fig.~\ref{fig:methods}. From left to right, the columns show H\,{\sc {i}} total intensity maps, H\,{\sc {i}} velocity fields, H\,{\sc {i}} velocity dispersion maps, distributions of H\,{\sc {i}} velocity dispersion, and integrated spectral profiles, respectively.  From top to bottom, the rows show the above-mentioned maps for the full H\,{\sc {i}} data cube, the regularly-rotating H\,{\sc {i}} component, and the anomalous H\,{\sc {i}} component. 
The total intensity maps clearly show how the regularly-rotating H\,{\sc {i}} component corresponds to the dominant morphological features of the galaxy (e.g., spiral arms), whereas the anomalous H\,{\sc {i}} component is much more uniformly distributed. The velocity fields show both components to be rotating, yet the amplitude of rotation is lower for the anomalous component. 
The H\,{\sc {i}} velocity dispersion results show how the two H\,{\sc {i}} components contrast with one another.

\subsection{Validation using a particle-based analysis}

\begin{figure}
	\centering
	\includegraphics[scale = 0.4]{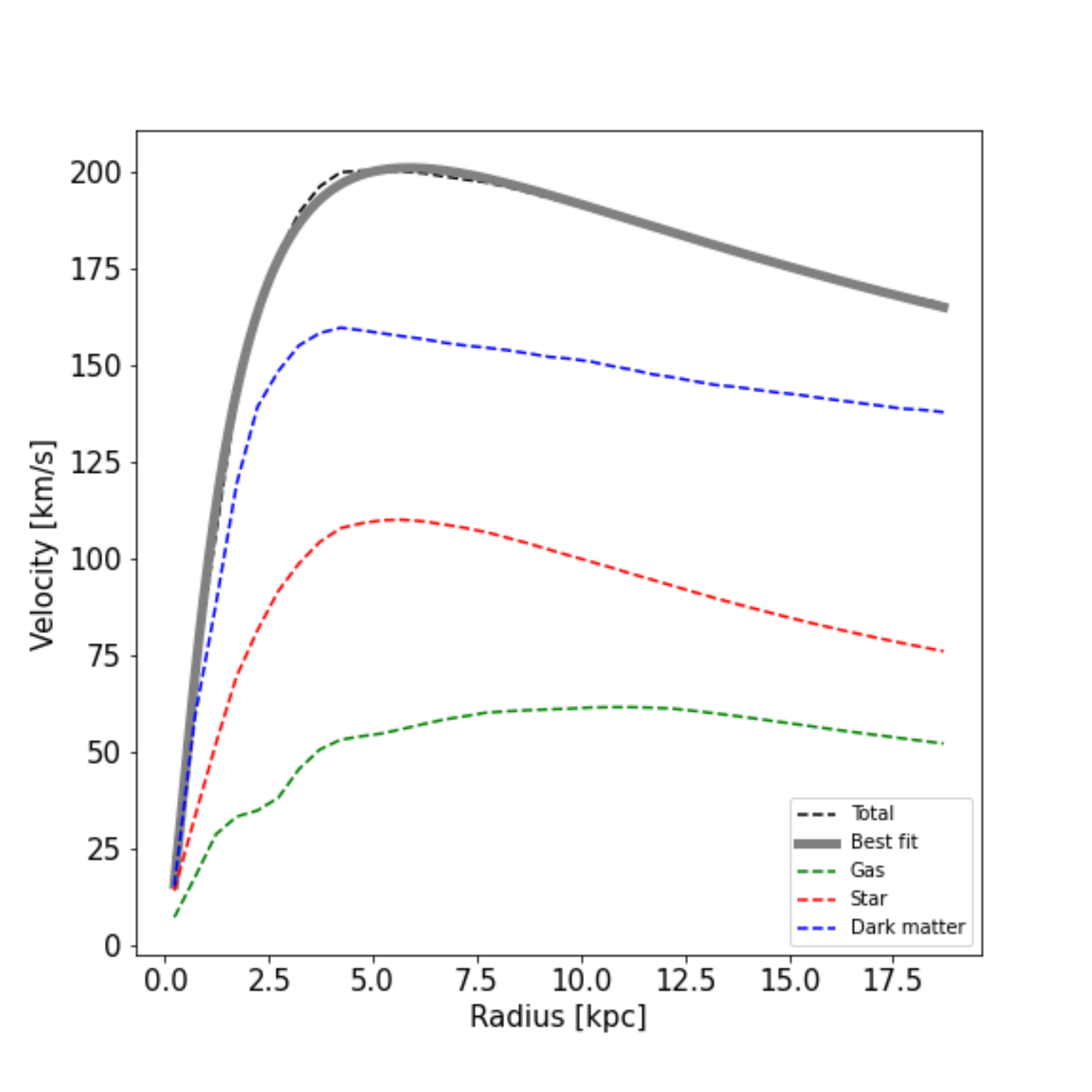}
	\caption{Measured rotation curve for each particle type in an example simulated galaxy, where the green, red, and blue dashed lines represent the stellar, gas, and dark matter contributions to the rotational velocity, respectively. The black dashed line is the total rotational velocity, while the grey curve represents the best fit to a theoretical profile.} 
	\label{fig:rotc}
\end{figure}

The decomposition method we describe in the previous subsection can be applied to simulated or observed H\,{\sc{i}} data cubes. Given our direct access to full phase space information (i.e., position and velocity vectors) for the particles making up each {\sc{Simba}} galaxy, we have also developed a second, particle-based decomposition method that 
we can use to separate the kinematically anomalous material from the main, regularly rotating disk, and thereby validate our cube-based method.
For each {\sc Simba} galaxy, we use CAESAR and pyGadgetReader \citep{soft11001T} to extract the position and velocity vectors of all the particles. For each set of particles (gas, stellar, and dark matter), we calculate the circular velocity of a test particle at radius $R$ due to all of the mass contained within a sphere of radius $R$. An idealised circular velocity curve of this nature is calculated for each mass component. The various curves are then added in quadrature to obtain the total circular velocity as a function of radius. Figure \ref{fig:rotc} shows the circular velocity curves generated for an example galaxy. To the total circular velocity curve, we fit the model from \cite{Wojnar_2018} that successfully fits observed data from H\,{\sc{i}} surveys like The HI Nearby Galaxy Survey (THINGS) \citep{de_Blok_2008,Walter_2008}:

\begin{equation}
v^2(r) = \frac{GM_0}{r} \left ( \sqrt{\frac{R_0}{r_c}}\frac{r}{r+r_c}\right )^{3\alpha} \left [ 1+ b\left ( 1+ \frac{r}{R_0}\right ) \right ]
\end{equation}
where $R_0$, $r_c$, $\alpha$, and $b$ are free parameters.  

\begin{figure*}
	\centering
	\includegraphics[scale= 0.35]{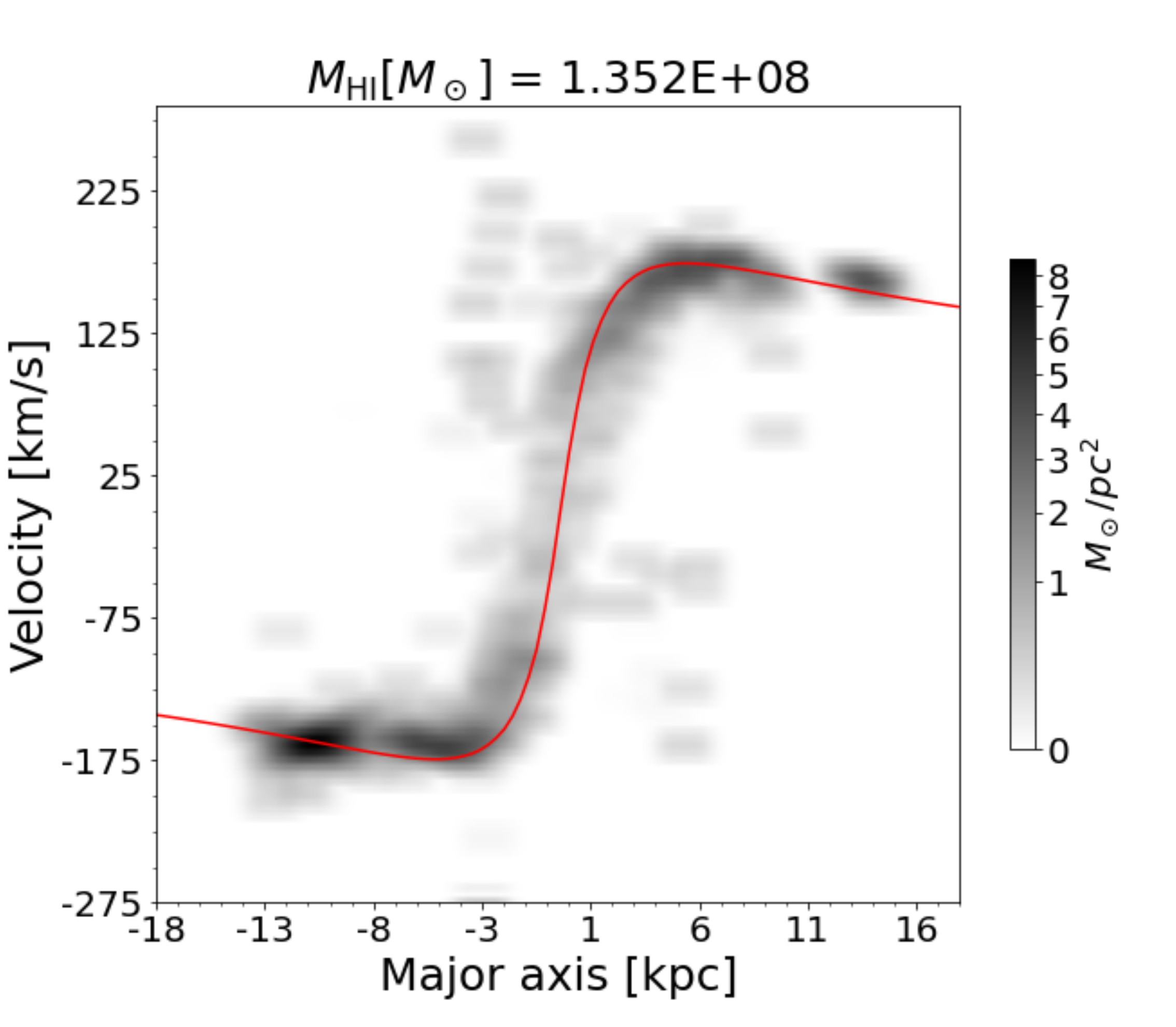}
	\includegraphics[scale = 0.35]{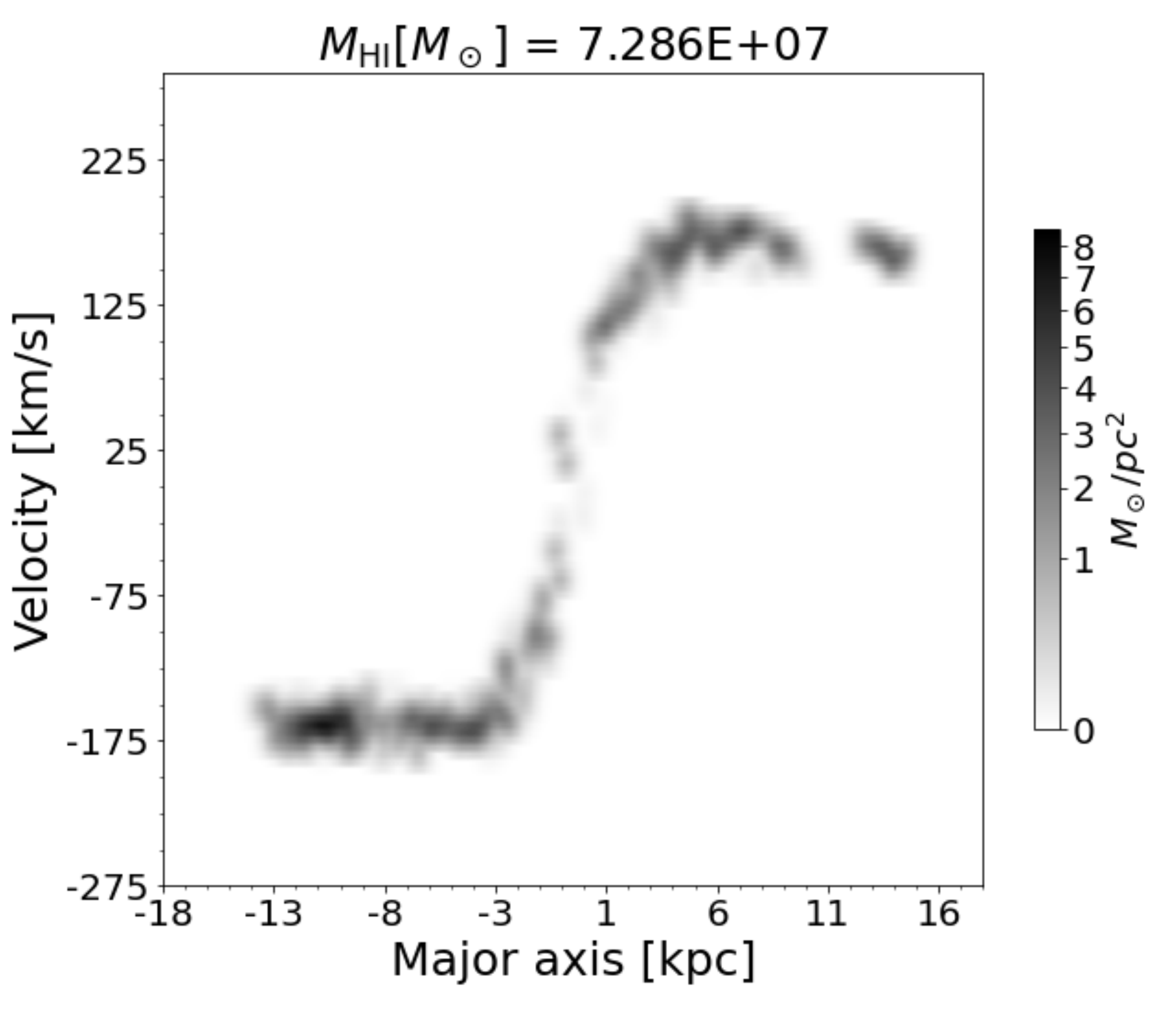}
	\caption{Left: Position-velocity diagram for the galaxy shown in Figure ~\ref{fig:rotc}. The red line is the expected rotation curve that comes from the best-fit rotation curve plotted in Figure ~\ref{fig:rotc}. The value on top is the total mass of H\,{\sc{i}} gas present in this slice. Right: Position-velocity diagram for the regularly rotating gas after removal of anomalous material using the particle-based decomposition method.}
	\label{fig:pvs}
	
\end{figure*}

Given the resulting model for the circular velocity as a function of radius, we split the particles of a given galaxy into spherical shells. In order to define the circular velocity component for each particle, we decompose the total velocity into two components: along the direction of the axis of rotation of the galaxy, and in the galactic plane. We then split the latter velocity component into radial and tangential components, where the radial unit vector is obtained from the projection of the particle's position vector onto the galactic plane. We then define the tangential velocity component as the particle's circular velocity component.
For each shell, we identify all particles that have their angular momentum vectors closely aligned to the total angular momentum vector of the galaxy (i.e., the angle between the momentum vectors is less than 45 deg) and have a velocity component tangential to the spherical shell that is within $30\,{\rm km\,s^{-1}}$ of the expected circular velocity (as predicted by the fitted rotation curve), to be the regularly-rotating subsample within the spherical shell, while the rest are taken to be kinematically anomalous. By using the idealised circular velocity curve to kinematically separate the particles on a shell-by-shell basis, we are able to generate entire data cubes containing the kinematically-regular and anomalous H\,{\sc{i}} components of {\sc Simba} galaxies. In left and right panels of Figure ~\ref{fig:pvs}, we show the position-velocity diagram of a galaxy before and after we apply this particle-based decomposition. The imperfect agreement between the rotation curve and the mock PV diagram is not surprising, since not all particles in a galaxy will be exactly in circular motion around its center of mass, and large scale asymmetries in a galaxy's gas distribution will not be captured in $v(r)$.

\begin{figure}
	\centering
	\includegraphics[scale = 0.45]{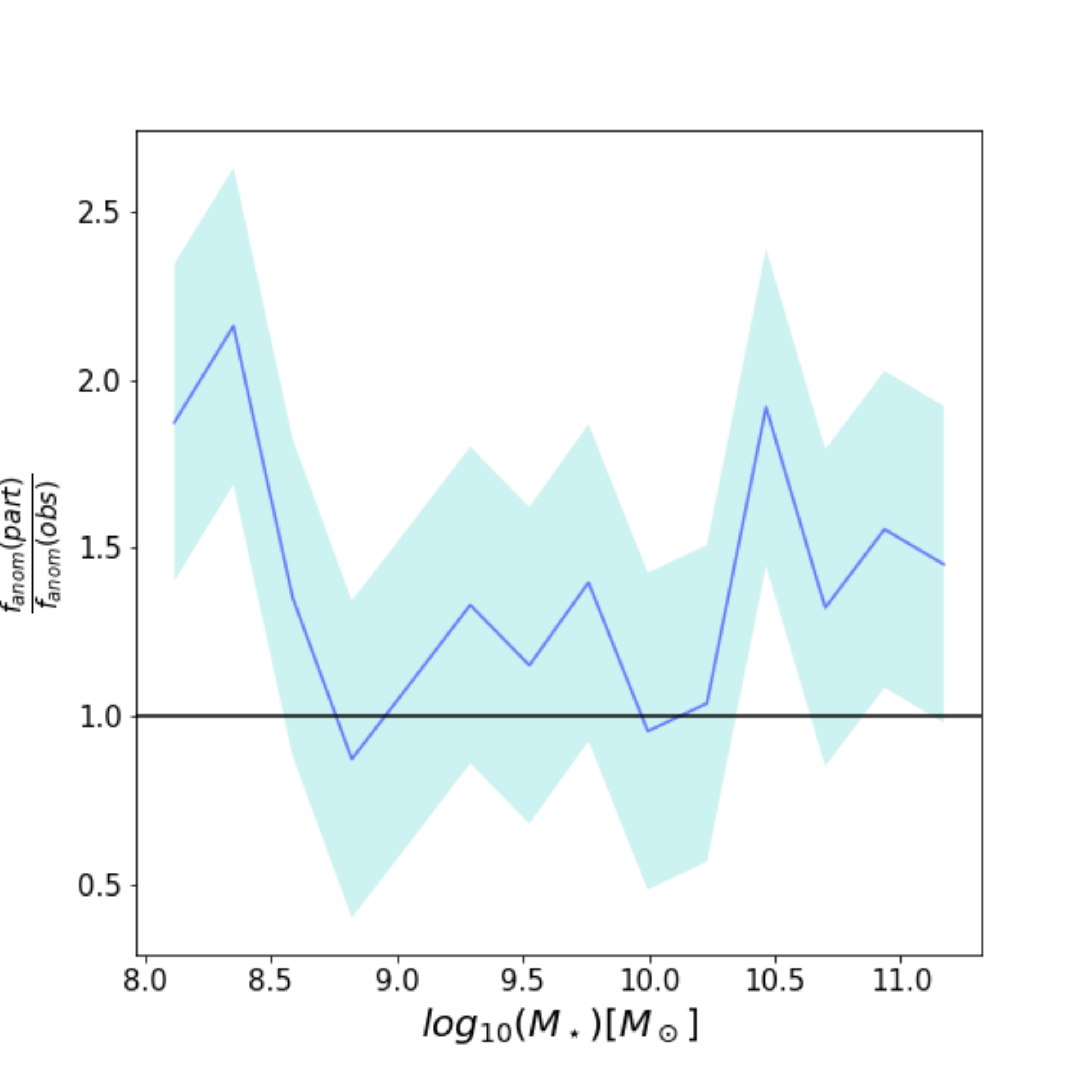}
	\caption{ Ratio of the particle-based anomalous gas fraction to the cube-based (``observational'') anomalous gas fraction as a function of stellar mass. The blue curve and cyan area represent the median and interquartile range of this ratio; the black line marks a ratio of unity.}
	\label{fig:comp}
\end{figure}

In Figure ~\ref{fig:comp}, over the range $8.5 < {\rm log}\,(M_\star/M_\odot) < 10$, the median ratio of anomalous gas fluxes determined using the cube-based and particle-based methods is close to unity. Below ${\rm log}\,(M_\star/M_\odot)=8.5$, the median ratio rises, possibly due to numerical and/or spatial resolution issues; the same increase also occurs at ${\rm log}\,(M_\star/M_\odot)>10.5$. However, over two orders of magnitude in stellar mass, the median ratio is less than a factor of 2, showing that the decomposition method we have devised for application to H\,{\sc{i}} data cubes should yield results that are reasonably consistent with those from a method that is applied directly to simulation particle data. Our cube-based method can therefore be reliably applied to observed H\,{\sc{i}} data cubes for galaxies with a wide range of stellar masses, in order to recover reliable measures of their anomalous H\,{\sc{i}} gas fractions.
An interesting (albeit computationally expensive) area for future work is investigating whether there is an inclination angle at which the particle-based and the cube-based methods would have approximately the same estimates of the anomalous fraction.

\subsection{Inclination effects}
	
\begin{figure*}

				\centering \includegraphics[scale = 0.45]{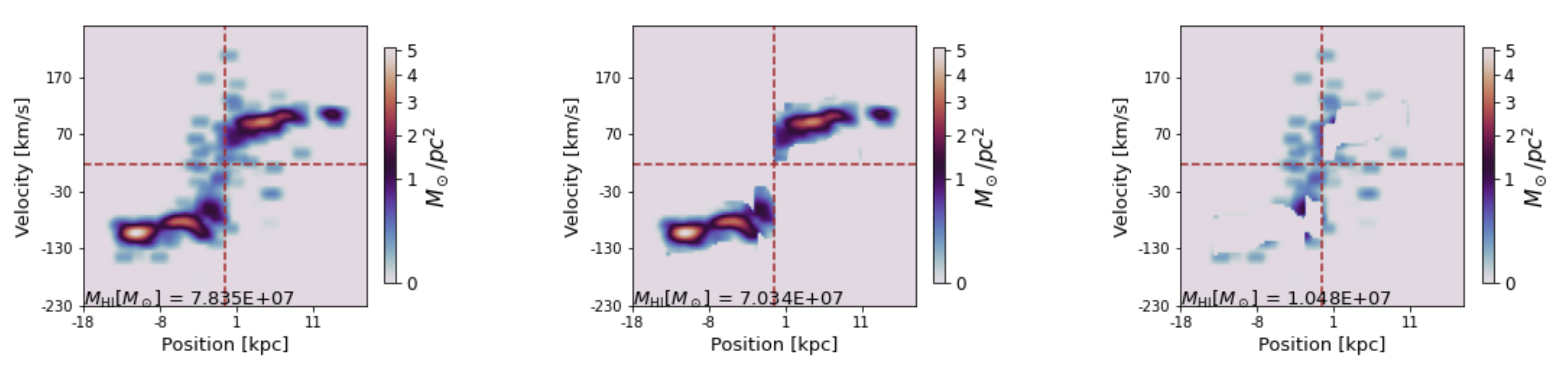} 
				\centering \includegraphics[scale = 0.45]{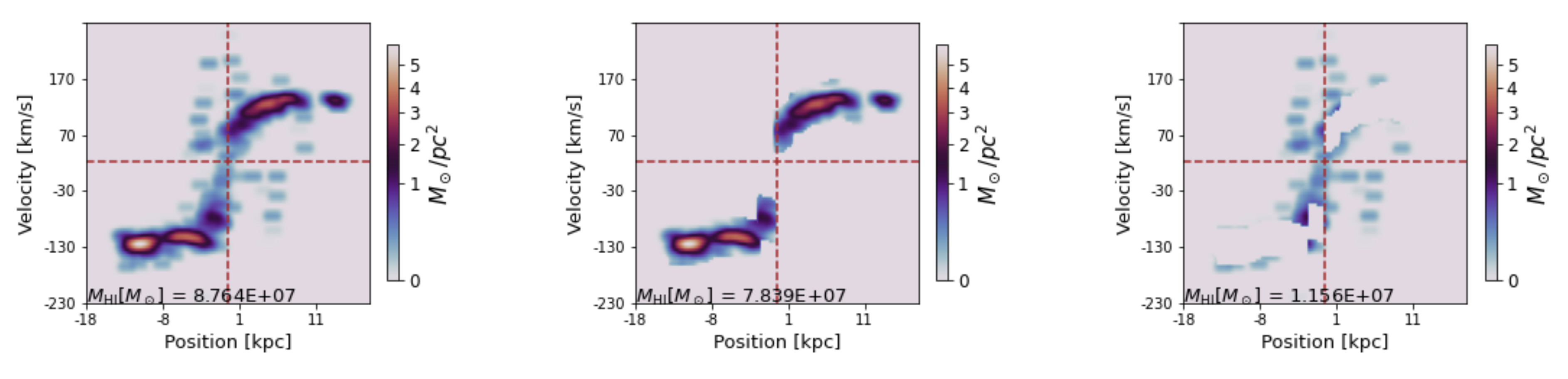} 
				\centering \includegraphics[scale = 0.45]{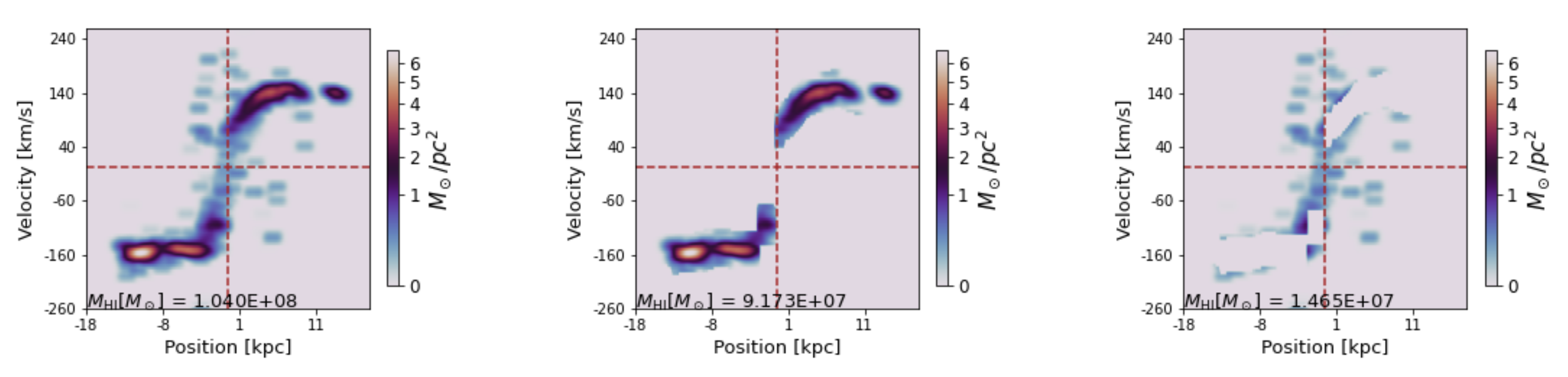} %
				\centering \includegraphics[scale = 0.14]{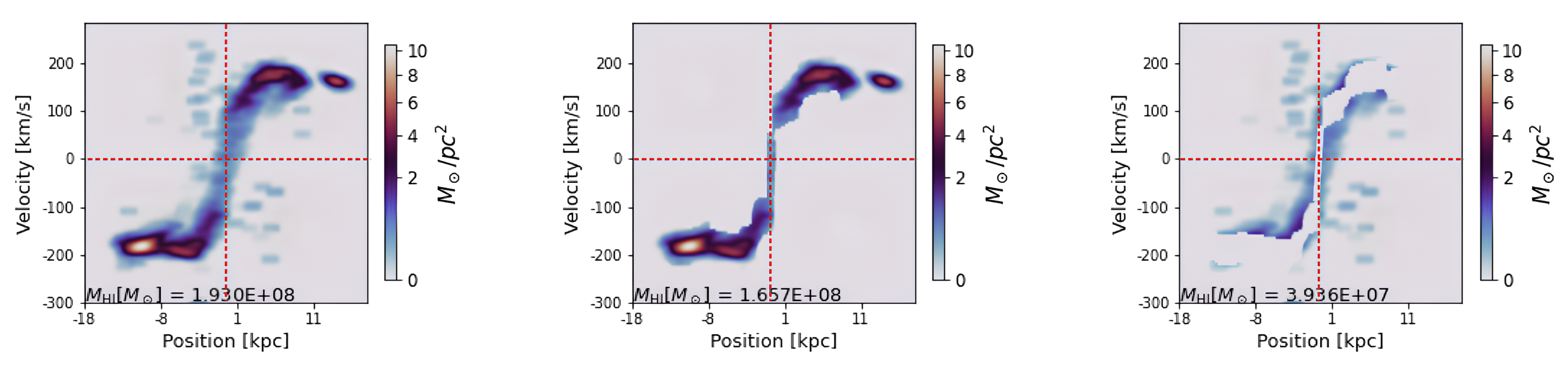} 
				\centering \includegraphics[scale = 0.45]{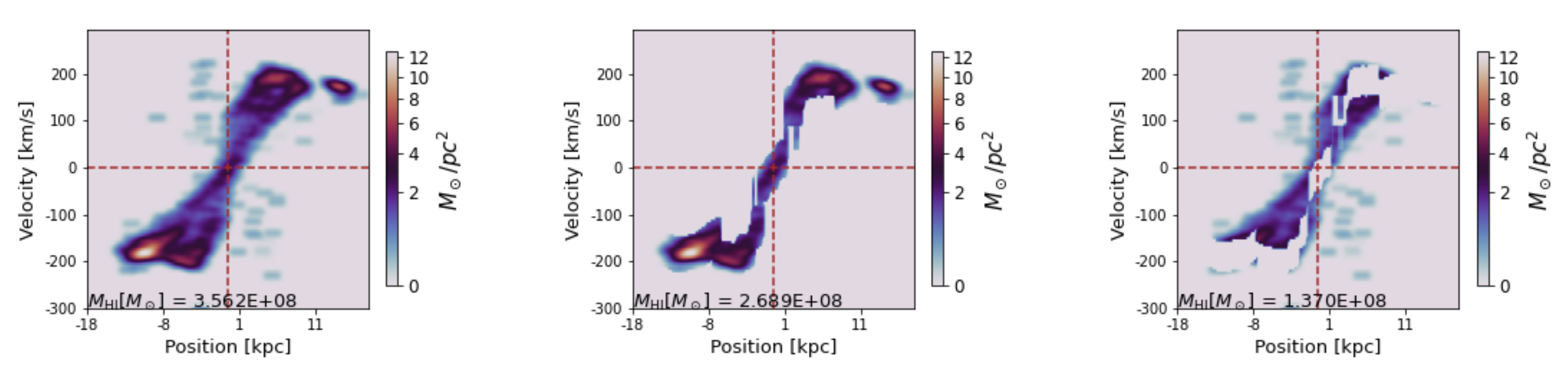}  
    
			\caption{Position-velocity (P-V) plots of the galaxy shown in Fig.~\ref{fig:PV1} at different inclination angles. Each row shows the P-V slice of the galaxy inclined at $30^\circ$, $40^\circ$, $50^\circ$, $70^\circ$, and $80^\circ$, respectively, from top to bottom. The first column comes from the full cube, while the second and third P-V slices are for the extracted regular and anomalous emission. The value shown on the upper left of each panel in the first column indicates the fraction of kinematically anomalous gas present in that slice. Values at the bottom indicate the total H\,{\sc {i}} mass present in the respective P-V slices.}
				\label{fig:PV30_75}
	\end{figure*}

\begin{figure*}
				
				\includegraphics[scale = 0.43]{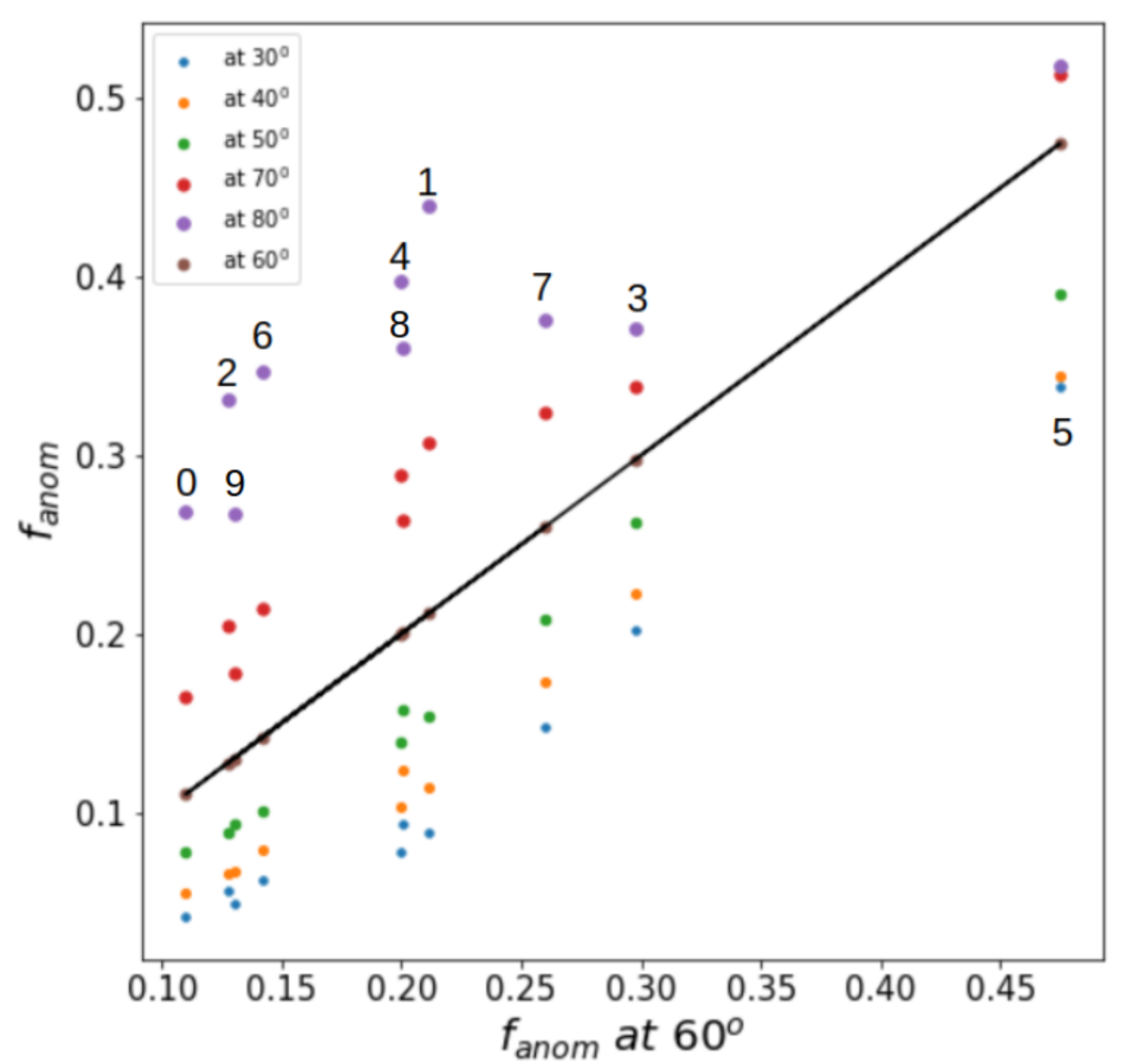}  				\caption{Anomalous gas fraction measured in ten simulated galaxies (numbered from 0 to 9) at different inclination angles, vs. the anomalous gas fraction measured at $60^\circ$ inclination angle. Each measurement of anomalous gas fraction is color coded differently for a different inclination angle from $30^\circ$ to $85^\circ$. The black line represents the one-to-one ratio of the anomalous gas fraction at $60^\circ$ inclination angle. Fig.~\ref{fig:panely} presents the moment 0 maps and H\,{\sc {i}} spectra of these galaxies. } 
				\label{fig:incli}
			\end{figure*}

\begin{figure*}
				
				\centering \includegraphics[scale = 0.22]{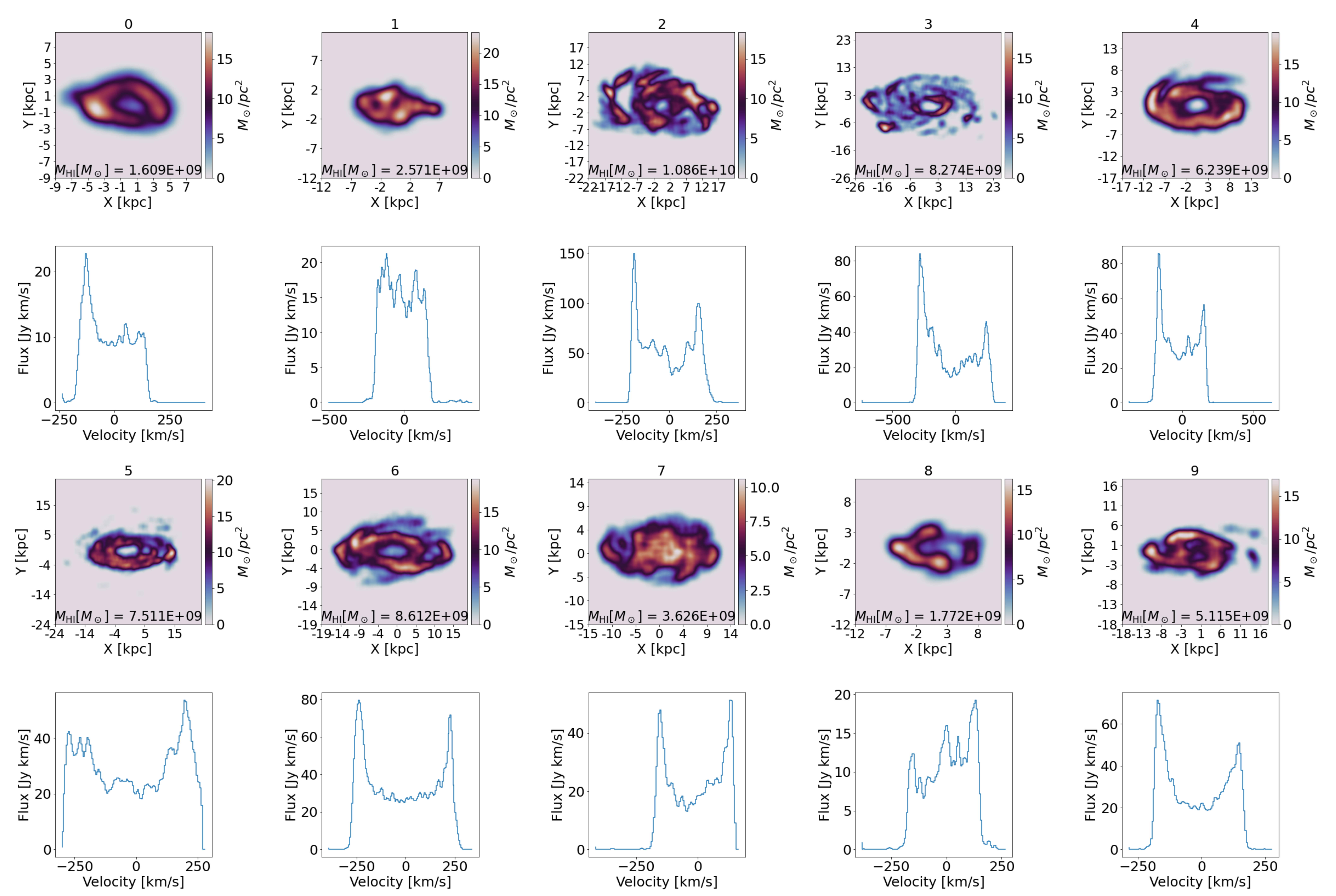}  
				\caption{Plots of the H\,{\sc {i}} moment 0 maps and profiles of the ten galaxies (numbered from 0 to 9) used to produce Fig.~\ref{fig:incli}}
				\label{fig:panely}
			\end{figure*}

At all spatial positions within a galaxy, the observed line-of-sight component of the total rotational motion of the gas is proportional to ${\rm sin}(i)$, where $i$ is the inclination of the disk. The more inclined the galaxy, the broader the spectrum, because the line of sight intercepts gas at a wider range of radii.
The shapes of observed line profiles are therefore affected by inclination, and therefore, the accuracy with which our method can decompose the kinematic components of a galaxy's gas will be affected by inclination. In Fig.~\ref{fig:PV30_75},

	we show P-V slices for the same galaxy shown in Fig.~\ref{fig:PV1}, but for versions of the H\,{\sc {i}} data cube that have the disk of the galaxy inclined at 30, 40, 50, 70, and 80 degrees (top to bottom) relative to the line of sight. For reference, the P-V slice shown in Fig.~\ref{fig:PV1} is based on a version of the H\,{\sc {i}} data cube that has the disk inclined at $60^\circ$. 
	We can see that the more edge-on a galaxy is, the larger the fraction of the total H\,{\sc {i}} mass our method identifies as anomalous.
	To further study the effect of inclination, we take ten large galaxies ($M_\star \ > 10^9 M_\odot$) from the simulation that are H\,{\sc {i}}-rich and disk-dominated and incline them at angles from $30^\circ$ to $80^\circ$ in steps of $10^\circ$. In Fig.~\ref{fig:incli}, 
	we show the recovered anomalous H\,{\sc {i}} gas fractions of the ten galaxies, numbered from 0 to 9 as indicated in their respective moment 0 maps and H\,{\sc {i}} profiles in Fig.~\ref{fig:panely}, vs. the \enquote{fiducial anomalous gas fraction}
	that we measure at $60^\circ$ inclination angle. We can see a clear dependence of the measured anomalous gas fraction on inclination angle, such that the more inclined a galaxy is, the higher its anomalous H\,{\sc {i}} content is. In fact, we can see that above 70 degrees, the anomalous gas fraction increases significantly. This analysis indicates that our decomposition method produces reliable results only for intermediate inclinations, and should not be applied to highly inclined systems in observational datasets. We also note that the quantitative inclination bias we measure here is connected to the specific parameter choices we have made when implementing our method, in particular the use of a $\pm 30\,{\rm km\,s^{-1}}$ velocity offset. Different parameter choices may ultimately lead to different ranges of inclination where our method can be reliably used.

	\section{Results and discussion} 
	\label{s-results}
	\subsection{Relationship between anomalous gas and star formation}
	
	The properties of neutral hydrogen in galaxies play a very important role in star formation, for which H\,{\sc {i}} serves as a major (albeit indirect) source of fuel. The location of galaxies with respect to the star-forming main sequence has been shown to be driven by gas fraction ~\citep{2016MNRAS.462.1749S, 2021arXiv210502413N}. However, the amount and nature of the gas in galaxies that feeds star formation is still poorly constrained. For the galactic fountain scenario, we would expect kinematically anomalous gas located in the halos of galaxies to be closely linked to the regions of highest star formation, such as spiral arms, which have been found to enhance star formation rates in disk galaxies ~\citep{2021arXiv210609715Y}. However, studies by ~\cite{2002AJ....123.3124F} and ~\cite{2005A&A...431...65B} of the anomalous gas in nearby galaxies show it is not always distributed like the observed spiral structures of the stellar disks. ~\cite{2019A&A...631A..50M} find only a weak correlation between star formation rate and the amount of kinematically anomalous gas in nearby galaxies. In this work, we use our newly-developed method of decomposing a galaxy's total H\,{\sc {i}} content into regularly rotating and kinematically anomalous components to search for links between these gas components and the star formation properties of the galaxies.\\ 
	
\begin{figure*}
		\includegraphics[scale = 0.56]{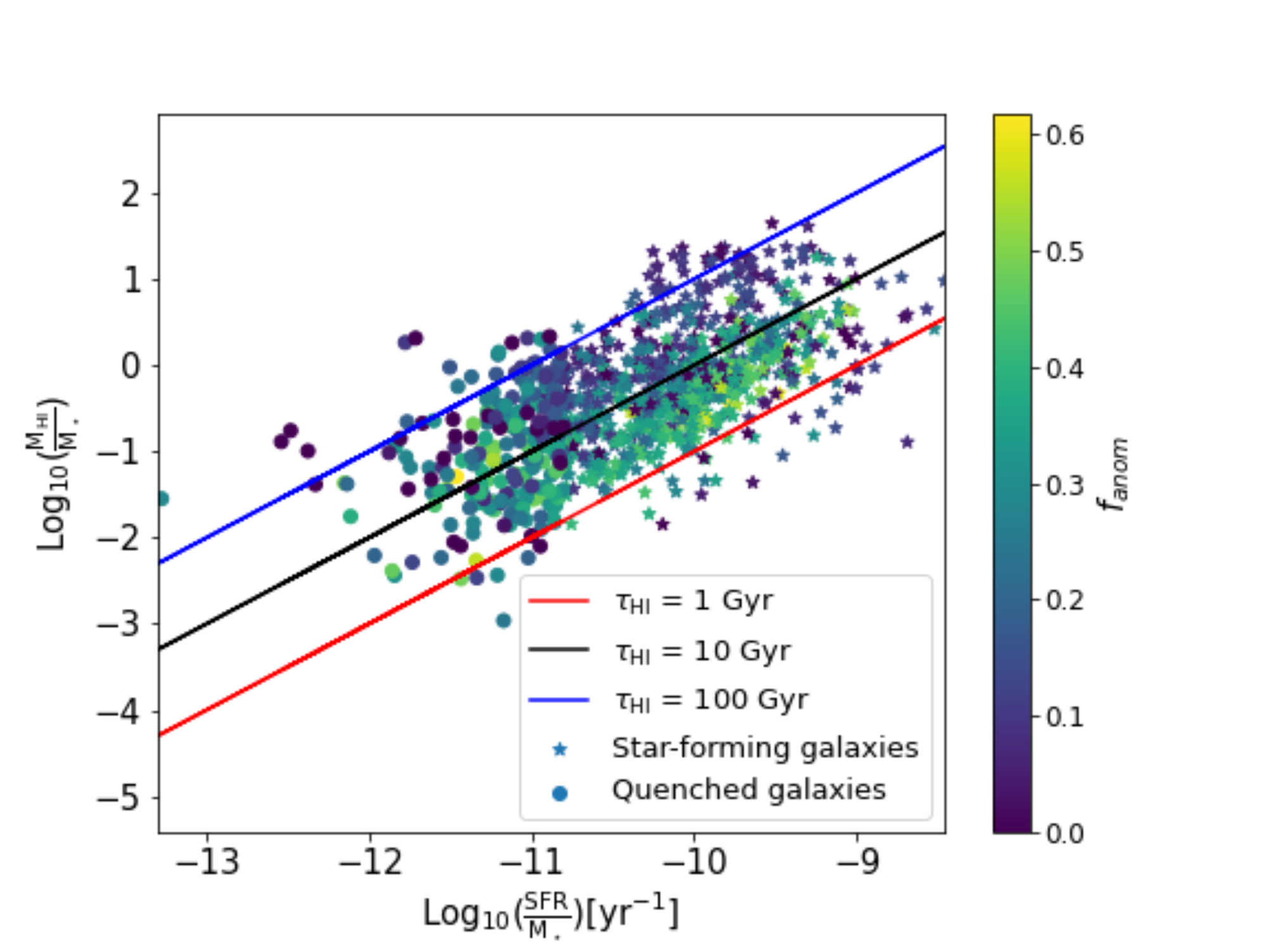}  
		\caption{Dependance of specific star formation rate on H\,{\sc {i}} gas fraction color-coded by the anomalous gas mass fraction. Star-forming galaxies are represented by star symbols, while quenched galaxies are represented by dot symbols. The three lines indicate atomic gas depletion times of 1, 10, and 100 Gyr (red, black, and blue respectively).}
		
		\label{fig:cor1}
		
	\end{figure*}

	In order to investigate the significance of the kinematically anomalous gas fraction, we show in Fig.~\ref{fig:cor1} the dependence of the specific star formation rate (${\rm sSFR} \equiv {\rm SFR}/M_\star$) on the atomic gas to stellar mass ratio ($\equiv M_{\rm H\,I}/M_\star$), color-coded by the anomalous gas mass fraction $f_{\rm anom}$ ($\equiv M_{{\rm H\,I},\rm anom}/(M_{{\rm H\,I},\rm anom}+M_{{\rm H\,I},\rm regular})$). 
	We overlay as solid lines H\,{\sc{i}} depletion times ($\tau_{\rm H\,I} \equiv M_{\rm H\,I}/{\rm SFR}$, the inverse of star formation efficiency) of 1 Gyr, 10 Gyr, and 100 Gyr colored red, black, and blue respectively. As expected, log\,(sSFR) increases with ${\rm log}\,(M_{\rm H\,I}/M_\star)$ for star-forming galaxies (star symbols), while quenched galaxies (circle symbols) by definition have low sSFR and are found strictly at low $M_{\rm H\,I}/M_\star$.  For the star-forming galaxies, $f_{\rm anom}$ clearly tends to {\it increase} with sSFR at fixed $M_{\rm H\,I}/M_\star$, while also slightly {\it decreasing} with $M_{\rm H\,I}/M_\star$ at fixed sSFR. The former trend can be explained if the kinematically anomalous gas boosts, or is boosted by, star formation in the disk. The latter trend also holds for the quenched galaxies (by definition, at lower fixed sSFR), suggesting that H\,{\sc{i}} kinematics are systematically more regular for galaxies where the H\,{\sc{i}} mass fraction is higher, independent of the level of star formation.  
	We can see also that $\tau_{\rm H\,I}$ decreases for increasing $f_{\rm anom}$. This result further constrains scenarios in which the anomalous gas is either a cause or an effect of enhanced sSFR. If the anomalous H\,{\sc{i}} is, e.g., infalling \enquote{fountain} material that enhances star formation, then it must do so in a way that accelerates its own consumption (or at least, conversion into ${\rm H_2}$). As a first step towards assessing causality here, we have  investigated the detailed histories of individual particles in a set of 20 galaxies, with five galaxies drawn at random from each of four stellar mass bins:
 8 $\leq {\rm log}(M_\star/M_\odot)$, 9 $\leq {\rm log}(M_\star/M_\odot) \leq 10$, 10 $\leq {\rm log}(M_\star/M_\odot) \leq 11$, and ${\rm log}(M_\star/M_\odot) > 11$. For all 20 galaxies, we applied our particle based method to separate the kinematically anomalous from the regularly rotating particles. We then traced the detected $z = 0$ anomalous particles back in time up to redshift $z = 0.2$, calculating the distance between each particle and the center of mass of its host galaxy at a given redshift. In most cases, we find the anomalous particles’ distances from their host galaxies’ centers decrease monotonically with time, suggesting that the kinematically anomalous particles are enhancing the star formation rate rather than being driven into an anomalous state by star formation feedback.

	\subsection{Relationship between anomalous gas and environment}
	
We can also investigate the possible impact of a galaxy's environment on its anomalous H\,{\sc {i}} gas content.~\cite{doi:10.1146/annurev.aa.22.090184.002305},~\cite{2006A&A...449..929G},~\cite{2009AN....330..904B}, and~\cite{2016MNRAS.461.2630M} find that at $z = 0$, the H\,{\sc {i}} morphology of galaxies are generally perturbed by ram pressure stripping, and that at fixed halo mass, the fraction of satellites devoid of H\,{\sc {i}} decreases as stellar mass increases due to the greater depth of the potential well. ~\cite{2020MNRAS.493.3238M} infer the loss of ionized gas in galaxies located in compact groups due to their environment, although the effect is small. However, a more recent study in ~\cite{2021arXiv210502413N} finds that H\,{\sc {i}} scaling relations are not driven by the environmental density per se.\\

\begin{figure*}
		\includegraphics[scale = 0.47]{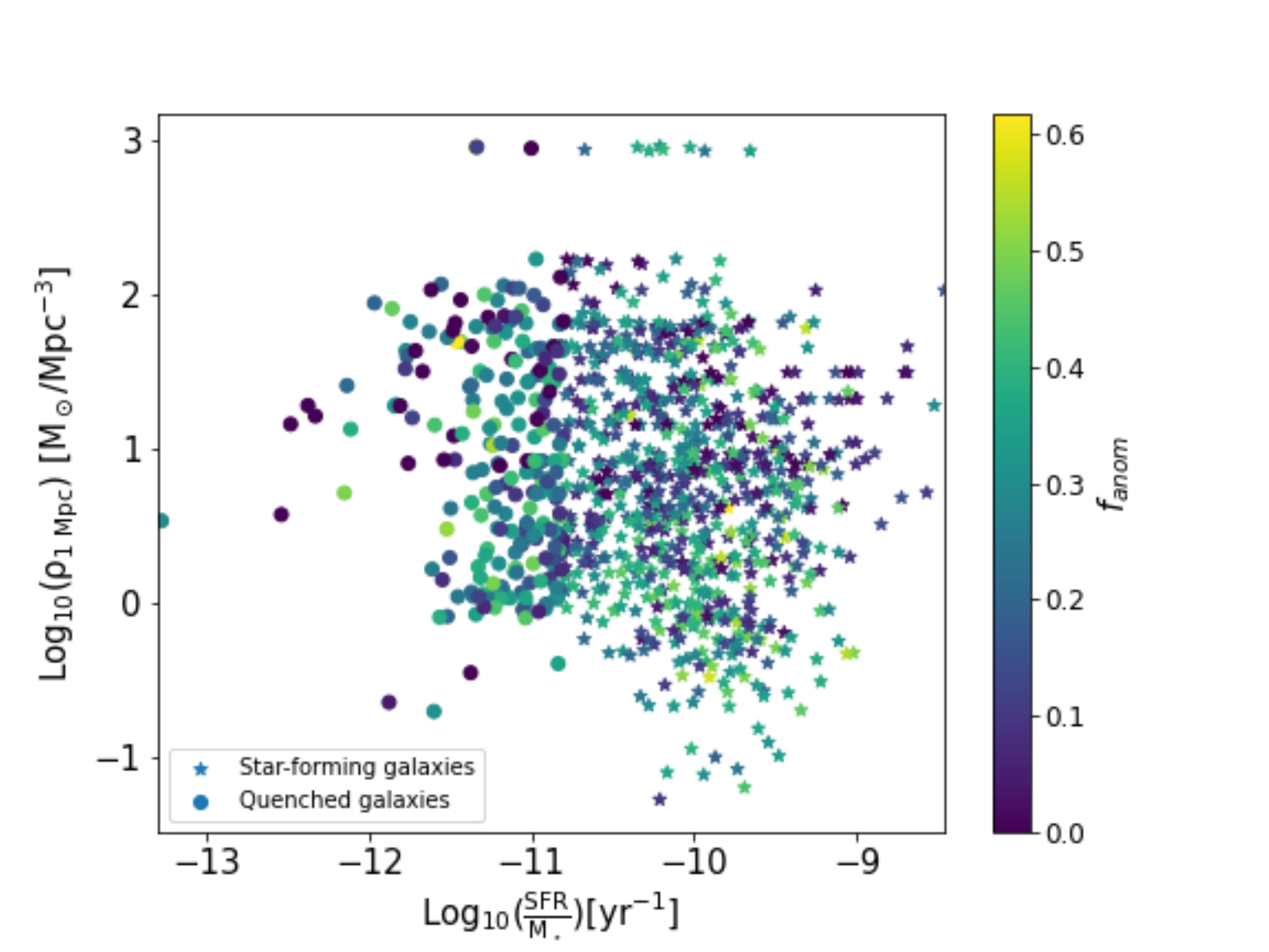}  
		\includegraphics[scale = 0.36]{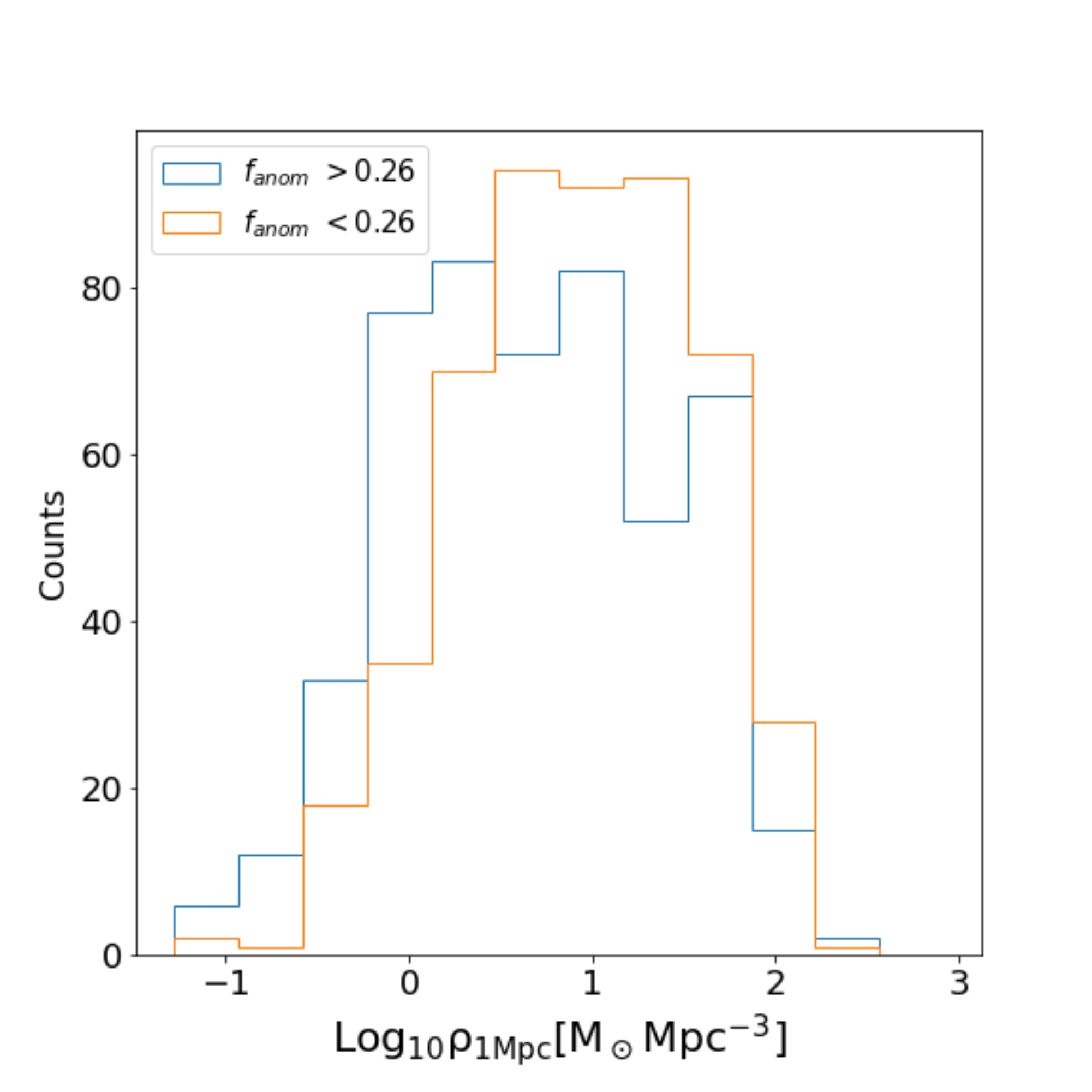}   
		\caption{Left: Mean stellar mass density in a sphere of radius 1Mpc ($\log (\rho_{1\rm Mpc})$) vs. specific star formation rate, color-coded by the anomalous gas fraction. Star-forming galaxies are represented by star symbols, while quenched galaxies are represented by dot symbols. Right: Distribution of the environmental density for galaxies with high and low anomalous gas fractions, represented by the orange and blue curves respectively.}
		\label{fig:rhom}
		
	\end{figure*}

In the left panel of Fig.~\ref{fig:rhom}, we show the relation between $\rho_{3{\rm Mpc}}$ the total stellar mass density averaged over a sphere of radius 1 Mpc (centered on each galaxy), and $\log({\rm sSFR})$. 
Points are again color-coded by the anomalous gas fraction. The highest $\log({\rm sSFR})$ are seen for dwarf galaxies, with $\log (\rho_{1\rm Mpc})$ decreasing as $\log({\rm sSFR})$ increases. 
This result is consistent with those seen in ~\cite{2011A&A...532A.145P} and ~\cite{2017MNRAS.466.2517M}. Dwarfs have low stellar mass yet high SFR due to the abundance of gas, and the mass density around dwarfs is lower than the mass density around more massive galaxies, since  massive galaxies tend to lie in richer environments. However, the anomalous gas fraction  $\log(f_{\rm anom})$ essentially depends on $\log({\rm sSFR})$ for both the star-forming and the quenched galaxies, as discussed above. To further show that environmental density does not have much impact on $\log(f_{\rm anom})$, we display in the right panel of Fig.~\ref{fig:rhom} a comparison of the distributions of $\rm log (\rho_{1\rm Mpc})$ for the galaxies that have high and low anomalous gas fractions (defined as above or below the median value  < $f_{\rm anom}$> = 26.47\%). We see clearly that the two distributions are not different from one another, with a Kolmogorov-Smirnov (K-S) test giving a p-value of 0.89 > 0.01, indicating that the two distributions are statistically identical. 
We therefore conclude that our measurement of the anomalous gas fraction is not as strongly dependent on environmental density as it is on star formation activity.	

\section{Conclusions} \label{s-conc}

In this work, we use the {\sc Simba} cosmological simulation to study the fraction of kinematically anomalous H\,{\sc {i}} in galaxies. We have developed a method that can be applied directly to a the H\,{\sc {i}} data cube of a galaxy in order to reliably decompose the total HI content into kinematically regular and anomalous components. We show the new cube-based method yields results similar to a method applied directly to the particle data from the simulation. We then discuss the dependence of the anomalous gas fraction on galaxy properties such as atomic hydrogen gas mass fraction, specific star formation rate, and environmental density. We find a significant effect of galaxy inclination on our measurement of the anomalous gas, which excludes it as a useful diagnostic for systems with $i\geq 70^\circ$. 
We find that anomalous gas fraction correlates well with specific star formation rate. This result can be interpreted as a sign that the anomalous H\,{\sc {i}} present in the galaxy enhances star formation, or that star formation pushes material out of the galactic disk and boosts the anomalous gas fraction. However, we do not see a significant change in the anomalous gas fraction as a function of galaxy environment, which may mean that kinematically anomalous H\,{\sc {i}} is mostly independent of the physical mechanisms that connect a host galaxy and its environment.    
In the near future, we plan to explore the application of our method to measure the anomalous gas fraction of observed galaxies in upcoming H\,{\sc {i}} surveys like LADUMA ~\citep{Blyth:2018RG} and MIGHTEE ~\citep{Taylor_2017}. 

\section*{Acknowledgments}

The authors thank the anonymous referee for feedback that significantly improved the paper.
NR thanks Marcin Glowacki for his insight on the simulations, and appreciates useful and helpful discussions with Amir Kazemi-Moridani, Anthony Young, and Michael Wozniak. NR acknowledges the use of computing facility of Inter-University Institute for Data Intensive Astronomy (IDIA) for this work. IDIA is a partnership of the University of Cape Town, the University of the Western Cape, and the University of Pretoria. EE acknowledges the support from the South African Radio Astronomy Observatory, which is a facility of the  National Research Foundation, an agency of the Departement of Sciences and Technology. This work is based on the research project supported wholly/in part by the National Research Foundation of South Africa (grant number 115238). AJB acknowledges support from the National Science Foundation via grant AST-1814421. The authors thank Robert Thompson for developing {\sc Caesar}, and \textsc{yt} team for the development and support of \textsc{yt}.

\bibliographystyle{mnras}
\bibliography{reference}

\begin{thebibliography}{}
\makeatletter
\relax
\def\mn@urlcharsother{\let\do\@makeother \do\$\do\&\do\#\do\^\do\_\do\%\do\~}
\def\mn@doi{\begingroup\mn@urlcharsother \@ifnextchar [ {\mn@doi@}
  {\mn@doi@[]}}
\def\mn@doi@[#1]#2{\def\@tempa{#1}\ifx\@tempa\@empty \href
  {http://dx.doi.org/#2} {doi:#2}\else \href {http://dx.doi.org/#2} {#1}\fi
  \endgroup}
\def\mn@eprint#1#2{\mn@eprint@#1:#2::\@nil}
\def\mn@eprint@arXiv#1{\href {http://arxiv.org/abs/#1} {{\tt arXiv:#1}}}
\def\mn@eprint@dblp#1{\href {http://dblp.uni-trier.de/rec/bibtex/#1.xml}
  {dblp:#1}}
\def\mn@eprint@#1:#2:#3:#4\@nil{\def\@tempa {#1}\def\@tempb {#2}\def\@tempc
  {#3}\ifx \@tempc \@empty \let \@tempc \@tempb \let \@tempb \@tempa \fi \ifx
  \@tempb \@empty \def\@tempb {arXiv}\fi \@ifundefined
  {mn@eprint@\@tempb}{\@tempb:\@tempc}{\expandafter \expandafter \csname
  mn@eprint@\@tempb\endcsname \expandafter{\@tempc}}}

\bibitem[\protect\citeauthoryear{Armillotta, Fraternali  \&
  Marinacci}{Armillotta et~al.}{2016}]{10.1093/mnras/stw1930}
Armillotta L.,  Fraternali F.,   Marinacci F.,  2016, \mn@doi [Monthly Notices
  of the Royal Astronomical Society] {10.1093/mnras/stw1930}, 462, 4157

\bibitem[\protect\citeauthoryear{{Barbieri}, {Fraternali}, {Oosterloo},
  {Bertin}, {Boomsma}  \& {Sancisi}}{{Barbieri}
  et~al.}{2005}]{2005A&A...439..947B}
{Barbieri} C.~V.,  {Fraternali} F.,  {Oosterloo} T.,  {Bertin} G.,  {Boomsma}
  R.,   {Sancisi} R.,  2005, \mn@doi [\aap] {10.1051/0004-6361:20042395}, \href
  {https://ui.adsabs.harvard.edu/abs/2005A&A...439..947B} {439, 947}

\bibitem[\protect\citeauthoryear{{Blyth} et~al.,}{{Blyth}
  et~al.}{2016}]{Blyth:2018RG}
{Blyth} S.,  et~al., 2016, in MeerKAT Science: On the Pathway to the SKA. p.~4

\bibitem[\protect\citeauthoryear{{Boomsma}, {Oosterloo}, {Fraternali}, {van der
  Hulst}  \& {Sancisi}}{{Boomsma} et~al.}{2005}]{2005A&A...431...65B}
{Boomsma} R.,  {Oosterloo} T.~A.,  {Fraternali} F.,  {van der Hulst} J.~M.,
  {Sancisi} R.,  2005, \mn@doi [\aap] {10.1051/0004-6361:20041715}, \href
  {https://ui.adsabs.harvard.edu/abs/2005A&A...431...65B} {431, 65}

\bibitem[\protect\citeauthoryear{{Boomsma}, {Oosterloo}, {Fraternali}, {van der
  Hulst}  \& {Sancisi}}{{Boomsma} et~al.}{2008}]{2008A&A...490..555B}
{Boomsma} R.,  {Oosterloo} T.~A.,  {Fraternali} F.,  {van der Hulst} J.~M.,
  {Sancisi} R.,  2008, \mn@doi [\aap] {10.1051/0004-6361:200810120}, \href
  {https://ui.adsabs.harvard.edu/abs/2008A&A...490..555B} {490, 555}

\bibitem[\protect\citeauthoryear{{Boselli}, {Boissier}, {Cortese}  \&
  {Gavazzi}}{{Boselli} et~al.}{2009}]{2009AN....330..904B}
{Boselli} A.,  {Boissier} S.,  {Cortese} L.,   {Gavazzi} G.,  2009, \mn@doi
  [Astronomische Nachrichten] {10.1002/asna.200911259}, \href
  {https://ui.adsabs.harvard.edu/abs/2009AN....330..904B} {330, 904}

\bibitem[\protect\citeauthoryear{{Bosma}}{{Bosma}}{1981}]{1981AJ.....86.1791B}
{Bosma} A.,  1981, \mn@doi [\aj] {10.1086/113062}, \href
  {https://ui.adsabs.harvard.edu/abs/1981AJ.....86.1791B} {86, 1791}

\bibitem[\protect\citeauthoryear{Dav{\'e}, Thompson  \& Hopkins}{Dav{\'e}
  et~al.}{2016}]{Dav__2016}
Dav{\'e} R.,  Thompson R.,   Hopkins P.~F.,  2016, \mn@doi [Monthly Notices of
  the Royal Astronomical Society] {10.1093/mnras/stw1862}, 462, 3265

\bibitem[\protect\citeauthoryear{Dav{\'e}, {\' e}s Alc{\' a}zar, Narayanan, Li,
  Rafieferantsoa  \& Appleby}{Dav{\'e} et~al.}{2019a}]{Dave:2019yyq}
Dav{\'e} R.,  {\' e}s Alc{\' a}zar D.,  Narayanan D.,  Li Q.,  Rafieferantsoa
  M.~H.,   Appleby S.,  2019a, \mn@doi [Mon. Not. Roy. Astron. Soc.]
  {10.1093/mnras/stz937}, 486, 2827

\bibitem[\protect\citeauthoryear{{Dav{\'e}}, {Angl{\'e}s-Alc{\'a}zar},
  {Narayanan}, {Li}, {Rafieferantsoa}  \& {Appleby}}{{Dav{\'e}}
  et~al.}{2019b}]{2019MNRAS.486.2827D}
{Dav{\'e}} R.,  {Angl{\'e}s-Alc{\'a}zar} D.,  {Narayanan} D.,  {Li} Q.,
  {Rafieferantsoa} M.~H.,   {Appleby} S.,  2019b, \mn@doi [\mnras]
  {10.1093/mnras/stz937}, \href
  {https://ui.adsabs.harvard.edu/abs/2019MNRAS.486.2827D} {486, 2827}

\bibitem[\protect\citeauthoryear{{Fraternali}, {Oosterloo}, {Sancisi}  \& {van
  Moorsel}}{{Fraternali} et~al.}{2001b}]{2001ApJ...562L..47F}
{Fraternali} F.,  {Oosterloo} T.,  {Sancisi} R.,   {van Moorsel} G.,  2001b,
  \mn@doi [\apjl] {10.1086/338102}, \href
  {https://ui.adsabs.harvard.edu/abs/2001ApJ...562L..47F} {562, L47}

\bibitem[\protect\citeauthoryear{Fraternali, Oosterloo, Sancisi  \& van
  Moorsel}{Fraternali et~al.}{2001a}]{Fraternali_2001}
Fraternali F.,  Oosterloo T.,  Sancisi R.,   van Moorsel G.,  2001a, \mn@doi
  [The Astrophysical Journal] {10.1086/338102}, 562, L47

\bibitem[\protect\citeauthoryear{{Fraternali}, {van Moorsel}, {Sancisi}  \&
  {Oosterloo}}{{Fraternali} et~al.}{2002}]{2002AJ....123.3124F}
{Fraternali} F.,  {van Moorsel} G.,  {Sancisi} R.,   {Oosterloo} T.,  2002,
  \mn@doi [\aj] {10.1086/340358}, \href
  {https://ui.adsabs.harvard.edu/abs/2002AJ....123.3124F} {123, 3124}

\bibitem[\protect\citeauthoryear{{Fraternali}, {Oosterloo}  \&
  {Sancisi}}{{Fraternali} et~al.}{2004}]{2004A&A...424..485F}
{Fraternali} F.,  {Oosterloo} T.,   {Sancisi} R.,  2004, \mn@doi [\aap]
  {10.1051/0004-6361:20040529}, \href
  {https://ui.adsabs.harvard.edu/abs/2004A&A...424..485F} {424, 485}

\bibitem[\protect\citeauthoryear{{Fraternali}, {Oosterloo}, {Sancisi}  \&
  {Swaters}}{{Fraternali} et~al.}{2005}]{2005ASPC..331..239F}
{Fraternali} F.,  {Oosterloo} T.~A.,  {Sancisi} R.,   {Swaters} R.,  2005, in
  {Braun} R.,  ed.,  Astronomical Society of the Pacific Conference Series Vol.
  331, Extra-Planar Gas. p.~239 (\mn@eprint {arXiv} {astro-ph/0410375})

\bibitem[\protect\citeauthoryear{{Gavazzi}, {O'Neil}, {Boselli}  \& {van
  Driel}}{{Gavazzi} et~al.}{2006}]{2006A&A...449..929G}
{Gavazzi} G.,  {O'Neil} K.,  {Boselli} A.,   {van Driel} W.,  2006, \mn@doi
  [\aap] {10.1051/0004-6361:20053844}, \href
  {https://ui.adsabs.harvard.edu/abs/2006A&A...449..929G} {449, 929}

\bibitem[\protect\citeauthoryear{{Haardt} \& {Madau}}{{Haardt} \&
  {Madau}}{2001}]{2001cghr.confE..64H}
{Haardt} F.,  {Madau} P.,  2001, in {Neumann} D.~M.,  {Tran} J.~T.~V.,  eds,
  Clusters of Galaxies and the High Redshift Universe Observed in X-rays. p.~64
  (\mn@eprint {arXiv} {astro-ph/0106018})

\bibitem[\protect\citeauthoryear{{Haffner} et~al.,}{{Haffner}
  et~al.}{2009}]{2009RvMP...81..969H}
{Haffner} L.~M.,  et~al., 2009, \mn@doi [Reviews of Modern Physics]
  {10.1103/RevModPhys.81.969}, \href
  {https://ui.adsabs.harvard.edu/abs/2009RvMP...81..969H} {81, 969}

\bibitem[\protect\citeauthoryear{Haynes, Giovanelli  \& Chincarini}{Haynes
  et~al.}{1984}]{doi:10.1146/annurev.aa.22.090184.002305}
Haynes M.~P.,  Giovanelli R.,   Chincarini G.~L.,  1984, \mn@doi [Annual Review
  of Astronomy and Astrophysics] {10.1146/annurev.aa.22.090184.002305}, 22, 445

\bibitem[\protect\citeauthoryear{Hopkins}{Hopkins}{2015}]{Hopkins_2015}
Hopkins P.~F.,  2015, \mn@doi [Monthly Notices of the Royal Astronomical
  Society] {10.1093/mnras/stv195}, 450, 53

\bibitem[\protect\citeauthoryear{{Kamphuis}, {Peletier}, {Dettmar}, {van der
  Hulst}, {van der Kruit}  \& {Allen}}{{Kamphuis}
  et~al.}{2007}]{2007A&A...468..951K}
{Kamphuis} P.,  {Peletier} R.~F.,  {Dettmar} R.~J.,  {van der Hulst} J.~M.,
  {van der Kruit} P.~C.,   {Allen} R.~J.,  2007, \mn@doi [\aap]
  {10.1051/0004-6361:20066989}, \href
  {https://ui.adsabs.harvard.edu/abs/2007A&A...468..951K} {468, 951}

\bibitem[\protect\citeauthoryear{Kannappan et~al.,}{Kannappan
  et~al.}{2013}]{Kannappan_2013}
Kannappan S.~J.,  et~al., 2013, \mn@doi [The Astrophysical Journal]
  {10.1088/0004-637x/777/1/42}, 777, 42

\bibitem[\protect\citeauthoryear{{Marasco}, {Crain}, {Schaye}, {Bah{\'e}}, {van
  der Hulst}, {Theuns}  \& {Bower}}{{Marasco}
  et~al.}{2016}]{2016MNRAS.461.2630M}
{Marasco} A.,  {Crain} R.~A.,  {Schaye} J.,  {Bah{\'e}} Y.~M.,  {van der Hulst}
  T.,  {Theuns} T.,   {Bower} R.~G.,  2016, \mn@doi [\mnras]
  {10.1093/mnras/stw1498}, \href
  {https://ui.adsabs.harvard.edu/abs/2016MNRAS.461.2630M} {461, 2630}

\bibitem[\protect\citeauthoryear{{Marasco} et~al.,}{{Marasco}
  et~al.}{2019}]{2019A&A...631A..50M}
{Marasco} A.,  et~al., 2019, \mn@doi [\aap] {10.1051/0004-6361/201936338},
  \href {https://ui.adsabs.harvard.edu/abs/2019A&A...631A..50M} {631, A50}

\bibitem[\protect\citeauthoryear{{Marinacci}, {Fraternali}, {Nipoti}, {Binney},
  {Ciotti}  \& {Londrillo}}{{Marinacci} et~al.}{2011}]{2011MNRAS.415.1534M}
{Marinacci} F.,  {Fraternali} F.,  {Nipoti} C.,  {Binney} J.,  {Ciotti} L.,
  {Londrillo} P.,  2011, \mn@doi [\mnras] {10.1111/j.1365-2966.2011.18810.x},
  \href {https://ui.adsabs.harvard.edu/abs/2011MNRAS.415.1534M} {415, 1534}

\bibitem[\protect\citeauthoryear{{Matsuki}, {Koyama}, {Nakagawa}  \&
  {Takita}}{{Matsuki} et~al.}{2017}]{2017MNRAS.466.2517M}
{Matsuki} Y.,  {Koyama} Y.,  {Nakagawa} T.,   {Takita} S.,  2017, \mn@doi
  [\mnras] {10.1093/mnras/stw2929}, \href
  {https://ui.adsabs.harvard.edu/abs/2017MNRAS.466.2517M} {466, 2517}

\bibitem[\protect\citeauthoryear{Meurer, Obreschkow, Wong, Zheng, Audcent-Ross
  \& Hanish}{Meurer et~al.}{2018}]{10.1093/mnras/sty275}
Meurer G.~R.,  Obreschkow D.,  Wong O.~I.,  Zheng Z.,  Audcent-Ross F.~M.,
  Hanish D.~J.,  2018, \mn@doi [Monthly Notices of the Royal Astronomical
  Society] {10.1093/mnras/sty275}, 476, 1624

\bibitem[\protect\citeauthoryear{{Miller} \& {Veilleux}}{{Miller} \&
  {Veilleux}}{2003}]{2003ApJS..148..383M}
{Miller} S.~T.,  {Veilleux} S.,  2003, \mn@doi [\apjs] {10.1086/376604}, \href
  {https://ui.adsabs.harvard.edu/abs/2003ApJS..148..383M} {148, 383}

\bibitem[\protect\citeauthoryear{{Moura} et~al.,}{{Moura}
  et~al.}{2020}]{2020MNRAS.493.3238M}
{Moura} T.~C.,  et~al., 2020, \mn@doi [\mnras] {10.1093/mnras/staa386}, \href
  {https://ui.adsabs.harvard.edu/abs/2020MNRAS.493.3238M} {493, 3238}

\bibitem[\protect\citeauthoryear{{Namiki}, {Koyama}, {Koyama}, {Yamashita},
  {Hayashi}, {Haynes}, {Shimakawa}  \& {Onodera}}{{Namiki}
  et~al.}{2021}]{2021arXiv210502413N}
{Namiki} S.~V.,  {Koyama} Y.,  {Koyama} S.,  {Yamashita} T.,  {Hayashi} M.,
  {Haynes} M.~P.,  {Shimakawa} R.,   {Onodera} M.,  2021, arXiv e-prints, \href
  {https://ui.adsabs.harvard.edu/abs/2021arXiv210502413N} {p. arXiv:2105.02413}

\bibitem[\protect\citeauthoryear{{Oosterloo}, {Fraternali}  \&
  {Sancisi}}{{Oosterloo} et~al.}{2007}]{2007AJ....134.1019O}
{Oosterloo} T.,  {Fraternali} F.,   {Sancisi} R.,  2007, \mn@doi [\aj]
  {10.1086/520332}, \href
  {https://ui.adsabs.harvard.edu/abs/2007AJ....134.1019O} {134, 1019}

\bibitem[\protect\citeauthoryear{{Popesso} et~al.,}{{Popesso}
  et~al.}{2011}]{2011A&A...532A.145P}
{Popesso} P.,  et~al., 2011, \mn@doi [\aap] {10.1051/0004-6361/201015672},
  \href {https://ui.adsabs.harvard.edu/abs/2011A&A...532A.145P} {532, A145}

\bibitem[\protect\citeauthoryear{{Popping}, {Dav{\'e}}, {Braun}  \&
  {Oppenheimer}}{{Popping} et~al.}{2009}]{2009A&A...504...15P}
{Popping} A.,  {Dav{\'e}} R.,  {Braun} R.,   {Oppenheimer} B.~D.,  2009,
  \mn@doi [\aap] {10.1051/0004-6361/200911811}, \href
  {https://ui.adsabs.harvard.edu/abs/2009A&A...504...15P} {504, 15}

\bibitem[\protect\citeauthoryear{{Rossa} \& {Dettmar}}{{Rossa} \&
  {Dettmar}}{2003}]{2003A&A...406..493R}
{Rossa} J.,  {Dettmar} R.~J.,  2003, \mn@doi [\aap]
  {10.1051/0004-6361:20030615}, \href
  {https://ui.adsabs.harvard.edu/abs/2003A&A...406..493R} {406, 493}

\bibitem[\protect\citeauthoryear{{Saintonge} et~al.,}{{Saintonge}
  et~al.}{2016}]{2016MNRAS.462.1749S}
{Saintonge} A.,  et~al., 2016, \mn@doi [\mnras] {10.1093/mnras/stw1715}, \href
  {https://ui.adsabs.harvard.edu/abs/2016MNRAS.462.1749S} {462, 1749}

\bibitem[\protect\citeauthoryear{Sales, Navarro, Theuns, Schaye, White, Frenk,
  Crain  \& Dalla~Vecchia}{Sales et~al.}{2012}]{20975}
Sales L.~V.,  Navarro J.~F.,  Theuns T.,  Schaye J.,  White S. D.~M.,  Frenk
  C.~S.,  Crain R.~A.,   Dalla~Vecchia C.,  2012, \mn@doi [Monthly Notices of
  the Royal Astronomical Society] {10.1111/j.1365-2966.2012.20975.x}, 423, 1544

\bibitem[\protect\citeauthoryear{{Schaap}, {Sancisi}  \& {Swaters}}{{Schaap}
  et~al.}{2000}]{2000A&A...356L..49S}
{Schaap} W.~E.,  {Sancisi} R.,   {Swaters} R.~A.,  2000, \aap, \href
  {https://ui.adsabs.harvard.edu/abs/2000A&A...356L..49S} {356, L49}

\bibitem[\protect\citeauthoryear{Swaters, Sancisi  \& van~der Hulst}{Swaters
  et~al.}{1997}]{Swaters_1997}
Swaters R.~A.,  Sancisi R.,   van~der Hulst J.~M.,  1997, \mn@doi [The
  Astrophysical Journal] {10.1086/304958}, 491, 140

\bibitem[\protect\citeauthoryear{Taylor \& Jarvis}{Taylor \&
  Jarvis}{2017}]{Taylor_2017}
Taylor A.~R.,  Jarvis M.,  2017, \mn@doi [{IOP} Conference Series: Materials
  Science and Engineering] {10.1088/1757-899x/198/1/012014}, 198, 012014

\bibitem[\protect\citeauthoryear{{Thompson}}{{Thompson}}{2014}]{soft11001T}
{Thompson} R.,  2014, {pyGadgetReader: GADGET snapshot reader for python},
  Astrophysics Source Code Library, record ascl:1411.001 (\mn@eprint {ascl}
  {1411.001})

\bibitem[\protect\citeauthoryear{{Verner} \& {Ferland}}{{Verner} \&
  {Ferland}}{1996}]{1996ApJS..103..467V}
{Verner} D.~A.,  {Ferland} G.~J.,  1996, \mn@doi [\apjs] {10.1086/192284},
  \href {https://ui.adsabs.harvard.edu/abs/1996ApJS..103..467V} {103, 467}

\bibitem[\protect\citeauthoryear{Walter, Brinks, de Blok, Bigiel, Kennicutt,
  Thornley  \& Leroy}{Walter et~al.}{2008}]{Walter_2008}
Walter F.,  Brinks E.,  de Blok W. J.~G.,  Bigiel F.,  Kennicutt R.~C.,
  Thornley M.~D.,   Leroy A.,  2008, \mn@doi [The Astronomical Journal]
  {10.1088/0004-6256/136/6/2563}, 136, 2563

\bibitem[\protect\citeauthoryear{Wojnar, Sporea  \& Borowiec}{Wojnar
  et~al.}{2018}]{Wojnar_2018}
Wojnar A.,  Sporea C.,   Borowiec A.,  2018, \mn@doi [Galaxies]
  {10.3390/galaxies6030070}, 6, 70

\bibitem[\protect\citeauthoryear{{Yu}, {Ho}  \& {Wang}}{{Yu}
  et~al.}{2021}]{2021arXiv210609715Y}
{Yu} S.-Y.,  {Ho} L.~C.,   {Wang} J.,  2021, arXiv e-prints, \href
  {https://ui.adsabs.harvard.edu/abs/2021arXiv210609715Y} {p. arXiv:2106.09715}

\bibitem[\protect\citeauthoryear{{Zschaechner}, {Rand}, {Heald}, {Gentile}  \&
  {Kamphuis}}{{Zschaechner} et~al.}{2011}]{2011ApJ...740...35Z}
{Zschaechner} L.~K.,  {Rand} R.~J.,  {Heald} G.~H.,  {Gentile} G.,   {Kamphuis}
  P.,  2011, \mn@doi [\apj] {10.1088/0004-637X/740/1/35}, \href
  {https://ui.adsabs.harvard.edu/abs/2011ApJ...740...35Z} {740, 35}

\bibitem[\protect\citeauthoryear{{de Blok}, {Walter}, {Brinks}, {Trachternach},
  {Oh}  \& {Kennicutt}}{{de Blok} et~al.}{2008}]{de_Blok_2008}
{de Blok} W.~J.~G.,  {Walter} F.,  {Brinks} E.,  {Trachternach} C.,  {Oh}
  S.-H.,   {Kennicutt} R.~C.,  2008, \mn@doi [The Astronomical Journal]
  {10.1088/0004-6256/136/6/2648}, 136, 2648

\makeatother
\end{thebibliography}
\label{lastpage}
\end{document}